\documentclass[aps,prx,showpacs,showkeys,twocolumn,superscriptaddress]{revtex4}
\usepackage{amssymb,amsmath}
\usepackage{bm,graphicx}
\usepackage{array}
\graphicspath{{Pictures/}}
\usepackage{natbib}
\usepackage{subfigure}
\usepackage{mathptmx}
\usepackage{pxfonts}
\usepackage{times}
\usepackage[T1]{fontenc}
\usepackage{epstopdf}
\usepackage[colorlinks,citecolor=blue,linkcolor=cyan]{hyperref}
\usepackage[usenames,dvipsnames,svgnames]{xcolor}

\begin{document}

\title{Granular Superconductor in a honeycomb lattice as a Realization of Bosonic Dirac Material}

\author{S. Banerjee}
\affiliation{Institute for Materials Sciences, Los Alamos National Laboratory, Los Alamos, New Mexico, 87545, USA}
\affiliation{Nordita, Center for Quantum Materials, KTH Royal Institute of Technology and Stockholm University,
Roslagstullsbacken 23, 10691 Stockholm, Sweden}
\affiliation{Division of Theoretical Chemistry and Biology, Royal Institute of Technology, SE-10691 Stockholm,Sweden}

\author{J. Fransson}
\affiliation{Department of Physics and Astronomy, Uppsala University, Box 516, S-751 20 Uppsala, Sweden}

\author{A. M. Black-Schaffer}
\affiliation{Department of Physics and Astronomy, Uppsala University, Box 516, S-751 20 Uppsala, Sweden}

\author{H. {\AA}gren}
\affiliation{Division of Theoretical Chemistry and Biology, Royal Institute of Technology, SE-10691 Stockholm,Sweden}

\author{A.V. Balatsky}
\affiliation{Institute for Materials Sciences, Los Alamos National Laboratory, Los Alamos, New Mexico, 87545, USA}
\affiliation{Nordita, Center for Quantum Materials,  KTH Royal Institute of Technology and Stockholm University, Roslagstullsbacken 23, 10691 Stockholm, Sweden}

\date{\today~  }
\begin{abstract}
We examine the low energy effective theory of phase oscillations in a two-dimensional granular superconducting sheet where the grains are arranged in honeycomb lattice structure. Using the example of graphene we present the evidence for the engineered Dirac nodes in the bosonic excitations: the spectra of the collective bosonic modes cross at the $K$ and $K'$ points in the Brillouin zone and form  Dirac nodes.    We show how two different types of collective phase oscillations are obtained and that they  are analogous to the Leggett and the Bogoliubov-Anderson-Gorkov modes in a two-band superconductor.  We show that the Dirac node is preserved in the presence of an inter-grain interaction, despite induced changes of the qualitative features of the two collective modes.  Finally, breaking the sublattice symmetry by choosing different on-site potentials for the two sublattices  leads to a gap opening near the Dirac node, in analogy with Fermionic Dirac materials. Dirac node dispersion of bosonic excitations is thus expanding the discussion of the conventional Dirac cone excitations to the case of bosons. We call this case as a representative of  Bosonic Dirac Materials (BDM), similar to the case of Fermionic Dirac materials extensively discussed in the literature.
\end{abstract}
\maketitle

\section{Introduction}

Over the last decade the honeycomb lattice has drawn significant attention within the condensed matter community. In addition to having interesting physical  properties the class of materials with this lattice structure also offers a realization of excitations with relativistic dispersion relation. In contrast to the conventional dispersion obtained by Schr\"odinger equation, these excitations are described by the Dirac equation. The most common example is graphene which exhibits massless low-energy Dirac fermions near the $K$ point in the Brillouin zone \cite{Castro}.  Other types of Dirac materials also exist, e.g., $d$-wave superconductors and surface states in topological insulators etc.~\cite{Wehling,Vafek}. These materials possess fermionic quasiparticle excitations, which can be described with linear Dirac-like energy-momentum dispersion relation for massless electrons. The crossing point of the bands in these materials are protected by different symmetries and breaking one of those symmetries leads to a gap opening near the Dirac point, e.g., in graphene a gap can be opened near the Dirac point by breaking the sublattice symmetry \cite{Semenoff,Park}.


\begin{figure}[t]
  \centering
  \includegraphics[width=.42\textwidth,trim= 0 80 0 0,clip ]{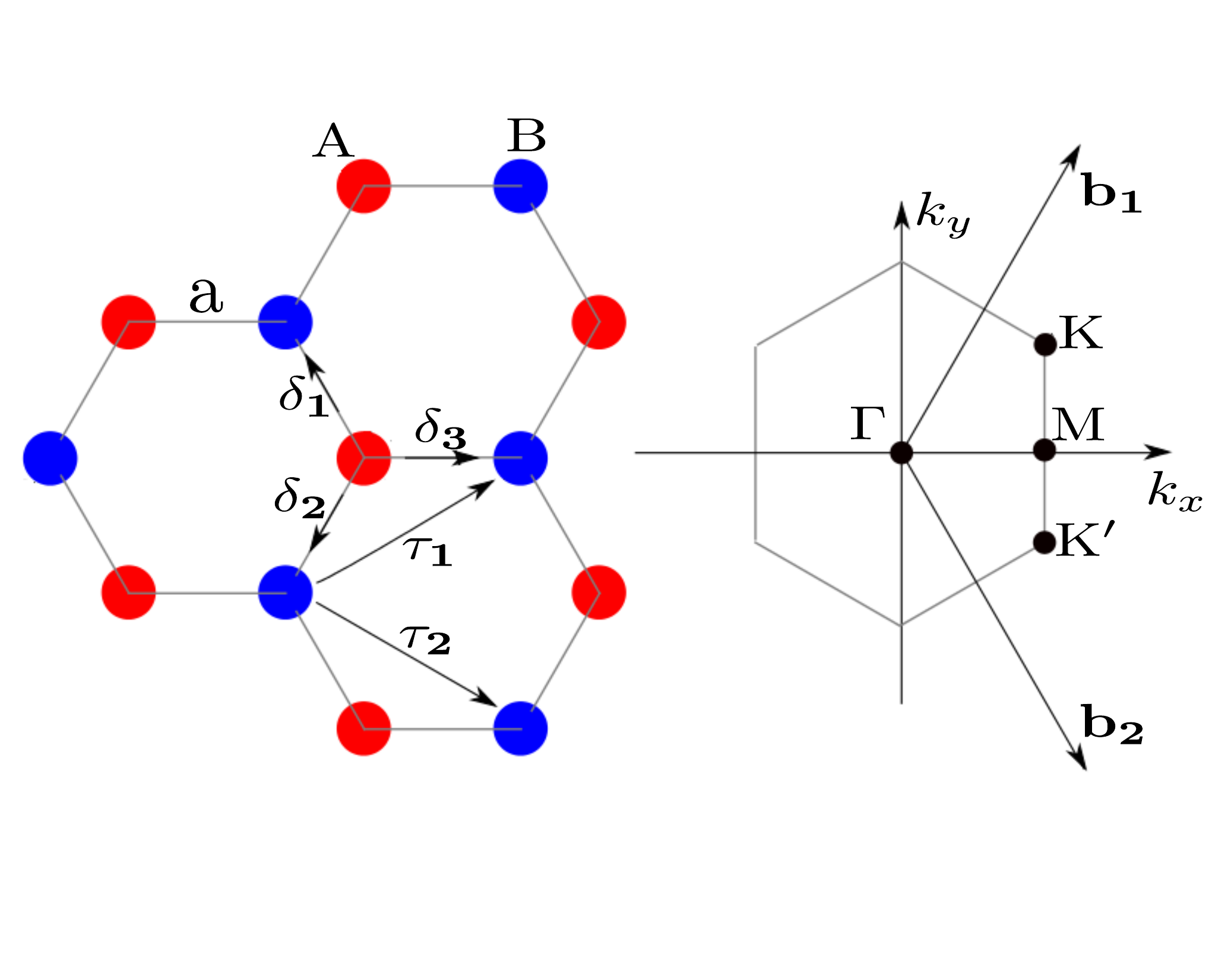}
  \caption{(Color online) Left: Lattice structure of superconducting grains (viz. Nb) built out of two triangular lattices with lattice vectors $\boldsymbol{\tau}_1$ and $\boldsymbol{\tau}_2$. The unit cell (red and blue dots) is composed of grains from two sublattices with nearest neighbor vectors $\boldsymbol{\delta}_{i}$, $i=1,2,3$. Adjacent grains interact by Josephson coupling ($J$). Individual grains have an on-site charging energy ($U$). Right: The corresponding Brillouin zone is shown, where ${\bf b}_1$ and ${\bf b}_2$ are the reciprocal lattice vectors, whereas $K$ and $K'$ are the Dirac points.}
  \label{fig:Honeycomb}
\end{figure}

From this point of view we can ask whether similar Bosonic Dirac materials (BDM) exists, in which the effective low energy quasiparticles are bosons. This question is motivated by the fact that the elemental carbon atoms in the graphene lattice have a bipartite lattice structure. Hence, despite each carbon atom being identical, the bipartite lattice structure ultimately leads to the Dirac equation in a tight-binding description of the carbon atoms in the graphene lattice \cite{Castro}. The logic for our analysis of BDM rests on the same observation, namely, that the single-particle hopping Hamiltonian of particles of any statistics between the nearest neighbors on a honeycomb lattice will have to generate Dirac nodes in the single-particle dispersion. Hence, one can envision experimental platforms that will generate the Dirac point in the excitation spectrum of bosons. To present the case of BDM and some of the universal features that are the consequence of Dirac node we will use the specific example of granular superconductors. While there are details that are relevant for the specific case of our model system we believe there are general statements that can be equally  applied to other materials with bosonic excitations that will have similar properties.   We point right away the key difference in dealing with bosons in comparison to fermions namely that interactions may have to be included for the emergence of non-trivial physics, for example a free 2D boson condensate in the ground state. Here, we discuss the potential to utilize Cooper pairs as effective bosons that occupy superconducting grains, to generate the BDM.

Recently we have seen the tremendous growth in our ability to manipulate the matter at the nano to mesoscale that allows us now to build structures with the desired properties like honeycomb lattices in materials and optical metamaterials. Here we mention some of the relevant work on artificial Dirac materials. Tarruell {\it et al.} studied the emergence and manipulation of Dirac nodes with cold atomic gases in a tunable honeycomb lattice ~\cite{Tarr}. In another important work authors have designed honeycomb lattice by individually moving carbon monoxide molecules with STM tip on a copper surface~\cite{Man1}. A detailed review~\cite{Man2} discussed about the recent progress on fabrication of artificial honeycomb lattice by different techniques like nano-patterning of two dimensional gases, STM guided molecule-by-molecule assembly or optical lattice. The work by Hammar {\it et al.}\cite{Hammar} is directly related to the experimental paper by Manoharan group \cite{Man1} and gives a proposal for artificial fabrication of lattice structures with tunable defects.   

 Recently, Weick {\it et al.} studied the collective plasmon modes in a honeycomb lattice, which showed Dirac-like massless bosonic excitations similar to Dirac fermions in graphene \cite{Weick}. Ref. \cite{Chen} numerically studies the non linear bloch bands in Bose-Einstein condensates (BEC) loaded in an optical honeycomb  lattice in the superfluid regime and discusses about band crossing between two of them. \cite{Wang} studies the Z$_{2}$ spin liquid states in honeycomb lattice in mean field theory and points out the resulting Dirac nodes in the dispersion. There is a growing interest in the study of photonic crystals and the electromagnetic waves around the sharp corners of the honeycomb lattice. Khanikaev {\it et al.} showed an example of Dirac-like dispersion in photonic topological insulators (PTI) ~\cite{Khan}. Carr {\it et al.} studied the emerging non-linear Dirac equation for BEC in optical honeycomb lattice ~\cite{Haddad}.  Propagation of microwaves in artificial honeycomb lattice composed of dielectric resonator and the resulting graphene like density of states were discussed in ref. \cite{Bellec}. Ref. \cite{Jacqmin} discusses about the observation of Dirac cones in the honeycomb lattice for polaritons . Schneider {\it et al.} discussed  the Dirac type Bloch band in BEC (Bose-Einstein condensate) ~\cite{Duca}. Ref. \cite{Sengupta} finds non-vanishing Berry curvature in honeycomb lattice of soft-core bosons and also talks about anomalous Hall effect in non-equilibrium. Saxena {\it et al.}  discussed  the Dirac cones in photonic crystal \cite{Chern}.  Dirac-like magnetic excitations were also pointed out for  magnon Dirac materials in Ref~\cite{Jonas}. All the above examples do prove that we can have artificial materials that exhibit Dirac-like dispersion for bosons in the solid state context, in cold atoms and in optical lattices are feasible and within the experimental reach similar.

 \begin{figure}[t]
  \centering
  \includegraphics[width=.42\textwidth]{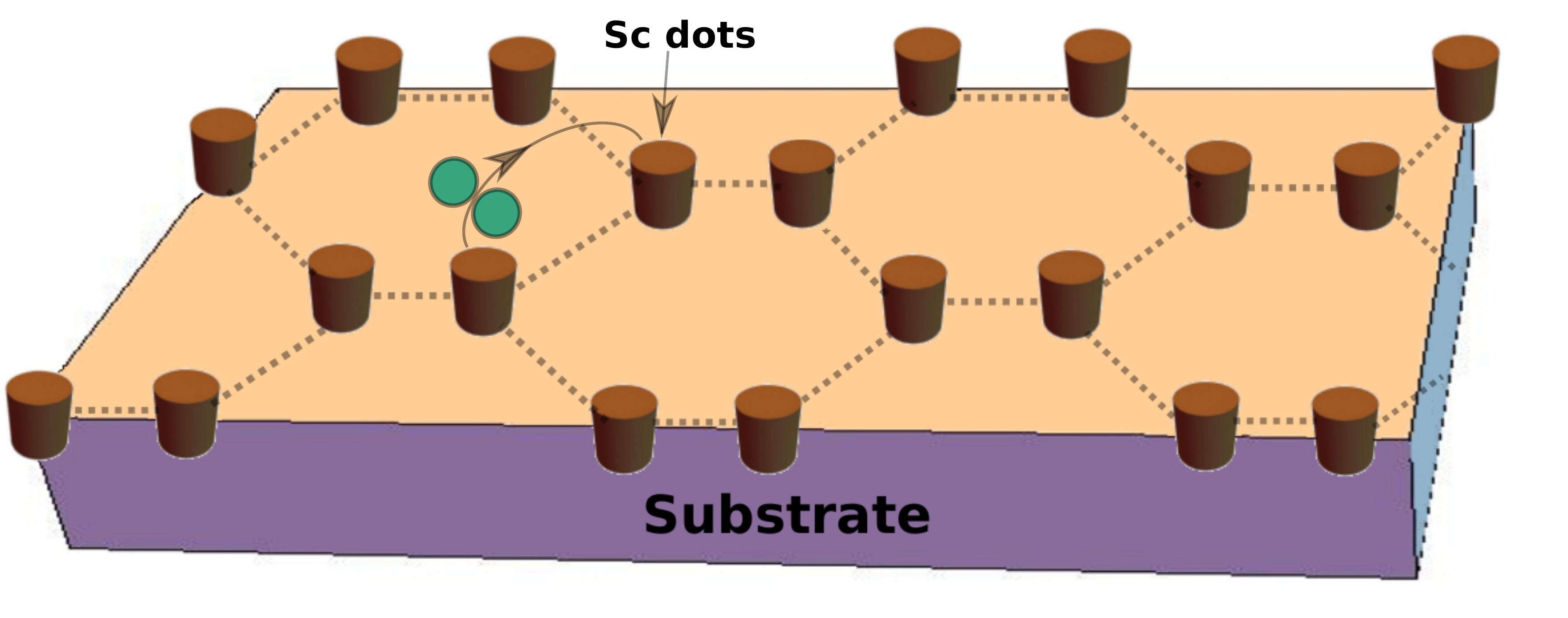}
  \caption{(Color online) A schematic of engineered quasi 2D hexagonal lattice structure of superconducting islands on an insulating substrate. The cooper pairs (shown in green) hop between the nearest neighbour islands. }
  \label{fig:Graphics}
\end{figure}

To develop the case for BDM we start with the specific example of granular  superconductors.  We will discuss the physics of the phase fluctuations of a granular superconducting system where the grains are arranged in a honeycomb lattice at temperature $T \ll T_c$, where $T_c$ is the mean-field superconducting transition temperature. The grains can be made of any conventional superconducting material and the choice depends on the practicality of sample preparations, see Fig.~\ref{fig:Graphics}  . The typical size (radius) of the grains is of the order of a few hundred to thousand of nanometers. Within the bipartite lattice structure, it is convenient to assign flavor, or, sublattice indices to the otherwise identical bosons.  The two types of bosons $b_A = |b^A|\exp(-i\theta^A)$ and $b_B = |b^B|\exp(-i\theta^B)$, have inequivalent phases $\theta^{A,B}$ on the grains (red dot $A$ and blue dot $B$ in Fig.~\ref{fig:Honeycomb}) in the unit cell of the lattice (we will assume that the amplitudes are same if all the grains are of approximately similar size). Nearest neighbor grains interact through a Josephson coupling $J$ and have a charging energy $U$ depending on the size of the grains, see Eq.~(\ref{eq:1}). We show that the physics of the inequivalent phases of the granular system is similar to the physics of a two-band superconductor and that there indeed are two different collective modes similar to the Leggett and Bogoliubov-Anderson-Gorkov (BAG) mode ~\cite{Belo} , also known in the literature as the Anderson-Higgs and Goldstone modes. We show that these two modes intersect and form a Dirac node near the $K$ and $K'$ points of the Brillouin zone in the bosonic excitation spectrum.

Interacting bosons in 2D have been the focus of intense investigation for some time especially the two dimensional granular superconducting systems \cite{Krull}. For temperatures below $T_c$, the grains support the existence of Cooper pairs. Therefore, at $T \ll T_c$ we can focus on the physics of the composite boson problem on a honeycomb lattice. Fisher {\it et al.} have studied this problem and shown in the clean boson limit that the system exhibits an interesting phase diagram showing 1) a superfluid phase and 2) a Mott insulating phase ~\cite{Gan}. In the disordered boson picture there is a bose glass phase \cite{Fisher,Kisker}. Recently, Doniach {\it et al.} proposed a new phase, called the Bose metal, in 2D superconducting films \cite{Das} in the presence of disorder. In this work, as we are building the case for the BDM, we will focus on the superfluid phase of the clean Bose-Hubbard model for honeycomb lattice.

The outline of this work is as follows. In Sec.~II we describe the formalism of the effective theory of granular superconductor. In Table~1 we summarize the main results of our work. In Sec.~II-A, we show the occurrence of the two collective modes Leggett and BAG. In Sec.~II-B we describe their low energy behavior of near $K$, $K'$ and $\Gamma$ point and show the of the Dirac-like behavior. In Sec.~III we describe the modification of the behavior of these collective modes by including the nearest neighbour interaction between the grains. In Sec.~III-A we describe their low energy behavior and show the Dirac cone. In Section IV, we discuss the tunability of the spectra and nodes at Dirac points and the role of disorder. Finally, we describe the outlook and conclusion of this work  in Sec.~V.

\section{Microscopic Hamiltonian for the granular superconductor: on-site interaction}
In this section we  describe an effective theory for the collective modes of phase oscillations in a 2D honeycomb lattice of superconducting grains (Fig. \ref{fig:Graphics}). We consider the Cooper pairs in each grain as charge 2e bosons, which are allowed to hop between the grains in the lattice. The lattice vectors are given by, $\boldsymbol{\tau}_1=a(3,\sqrt{3})/2$ and $\boldsymbol{\tau}_2=a(3,-\sqrt{3})/2$, where the lattice constant $a$ is of the order of a few to a few hundreds of $\mu$m. The reciprocal lattice vectors are given by ${\bf b}_1=2\pi(1,\sqrt{3})/3a$ and ${\bf b}_2=2\pi(1,-\sqrt{3})/3a$. Three nearest neighbor vectors are denoted by $\boldsymbol{\delta}_{1}=a(1,\sqrt{3})/2$, $\boldsymbol{\delta}_{2}=a(1,-\sqrt{3})/2$  and $\boldsymbol{\delta}_{3}=a(-1,0)$, see Fig.~\ref{fig:Honeycomb}. It is important to note that the grains are deposited on a substrate to form the quasi 2D lattice system.

\begin{table*}[t]
  \centering
  \caption[font=large]{$\omega_{1}({\bf k})$ is the Leggett mode frequency and $\omega_{2}({\bf k})$ is the BAG mode frequency. This Table contains the main results of the work. Cons. means a constant here. We use Cons. and Cons$'$. to differentiate between the two constants.}
  \begin{tabular}{| >{\centering\arraybackslash}m{4cm} | >{\centering\arraybackslash}m{6cm} | >{\centering\arraybackslash}m{6cm}| >{\centering\arraybackslash}m{1.5cm}|}
    \hline \hline
    \textbf{Parameter} & \textbf{$\Gamma$ Point} & \textbf{$K$ Point} &\textbf{Gap at K}\\
    \hline
    Free boson &  $\omega_{1}(\bf{k}) \approx Cons.-|\bf{k}|^2$ & $\omega_{1}(\bf{k}) \approx Cons. +|\bf{k}|$   & No\\
    $U_A=U_B=U_{AB}=0$ & $\omega_{2}(\bf{k}) \approx |\bf{k}|^2$ & $\omega_{2}(\bf{k}) \approx Cons.-|\bf{k}|$ & No\\

    \hline
    On-site Coulomb & $\omega^{2}_{1}(\bf{k}) \approx Cons.-|\bf{k}|^2$  & $\omega_{1}(\bf{k}) \approx Cons. + |\bf{k}|$ & No\\
    $U_A=U_B\neq 0$ $U_{AB}=0$ & $\omega_{2}(\bf{k}) \approx |\bf{k}|$ &  $\omega_{2}(\bf{k}) \approx Cons. - |\bf{k}|$ & No\\
    \hline
    On-site Coulomb & $\omega^{2}_{1}(\bf{k}) \approx Cons.-|\bf{k}|^{2}$  & $\omega^{2}_{1}(\bf{k}) \approx Cons. + |\bf{k}|^{2}$ & Yes\\
    $U_A\neq U_B\neq 0$ $U_{AB}=0$ & $\omega^{2}_{2}(\bf{k}) \approx Cons'. +|\bf{k}|^{2}$ &  $\omega_{2}^{2}(\bf{k}) \approx Cons'.- |\bf{k}|^{2}$ & Yes\\
    \hline
    Interacting Grains & $\omega_{1}^{2}(\bf{k}) \approx Cons. -|\bf{k}|^2$ & $\omega_{1}(\bf{k}) \approx Cons. + |\bf{k}|$ & No\\
    $U_A=U_B\neq 0$ $U_{AB}\neq0$ & $\omega_{2}^{2}(\bf{k}) \approx |\bf{k}|^2$ & $\omega_{2}(\bf{k}) \approx Cons. - |\bf{k}|$ & No\\
    \hline
    Interacting Grains & $\omega_{1}^{2}(\bf{k}) \approx Cons. -|\bf{k}|^2$ & $\omega^{2}_{1}(\bf{k}) \approx Cons. + |\bf{k}|^{2}$ & Yes\\
    $U_A \neq U_B\neq 0$ $U_{AB}\neq0$ & $\omega_{2}^{2}(\bf{k}) \approx Cons'.+|\bf{k}|^2$ &  $\omega^{2}_{2}(\bf{k}) \approx Cons'. - |\bf{k}|^{2}$ & Yes\\
    \hline
  \end{tabular}
\end{table*}

A Bose-Hubbard model can be written for this system by defining Cooper pair creation (annihilation) operators $b^{\dagger \alpha}_{i}=c^{\dagger \alpha}_{{\bf R}_{i}\uparrow}c^{\dagger \alpha}_{{\bf R}_{i}\downarrow}$ ($b^{\alpha}_{i}=c^{\alpha}_{{\bf R}_{i}\downarrow}c^{\alpha}_{{\bf R}_{i}\uparrow}$) in each grain, where $\alpha=A,B$ assigns to which sublattice the grain belongs. Here also, ${\bf R}_{i}$ denotes the spatial coordinate of the electrons in the i'th grain and is defined within a single granular size. Considering all the physics discussed in the introduction, we write down the Bose-Hubbard model in honeycomb lattice as
\begin{equation}\label{eq:1}
 \begin{aligned}
\mathcal{H}&=-\sum_{\langle ij \rangle}t_{ij}b^{\dagger A}_{i}b^{B}_{j} + h.c.+U\sum_{i,\alpha}(n^{\alpha}_{i}-n_{0})^{2}
,
\end{aligned}
\end{equation}

In Eq.~(\ref{eq:1}), $t_{ij}$ is the boson (Cooper pair) hopping amplitude and U is the on-site (Coulomb) charging energy for the bosons. The notation $\langle ij \rangle$ refers to the nearest neighbour hopping, $n_{0}$ is the neutralizing background density which is a large number such that long range Coulomb interactions can be avoided. We map the Bose-Hubbard model approximately to a quantum rotor model in the superfluid phase, by redefining the bose operators into a charge-density representation according to $b^{\dagger A}_{i} =\sqrt{n^{A}_{i}}\exp[i\theta^{A}_{i}]$ and $b^{A}_{i} =\exp[-i\theta^{A}_{i}] \sqrt{n^{A}_{i}}$. The operator $\exp[i\theta^{A}_{i}]$ denotes the Cooper pair creation operator whereas $\theta^{A}_{i}$ is the conjugate variable to the Cooper pair number operator $n^{A}_{i}$, which can be proved from the commutation relations of $[b_{i},b^{\dagger}_{j}]=\delta_{ij}$. As we are interested in the effective theory of the phase fluctuations in the superfluid phase, we drop the amplitude fluctuation in the bose operators and replace them by large value $n_{0}$ in Eq.~(\ref{eq:1}).

We examine the effective theory of phase fluctuations~\cite{Cha} from the following quantum rotor model, see Appendix A Eq.~(\ref{eq:27}),
\begin{equation}\label{eq:2}
\mathcal{H_{QR}}=-2J\sum_{\langle ij \rangle }\cos(\theta^{A}_{i}-\theta^{B}_{j})+ U\sum_{i\alpha}(n^{\alpha}_{i}-n_{0})^{2}
,
\end{equation}
In Eq.~(\ref{eq:2}), the Josephson coupling $J \sim n_{0}t$, where the nearest neighbor hopping is assumed to be uniform, $t_{ij}=t$, for all neighbors. By shifting $n_{i}\rightarrow n_{i}+n_{0}$, we can rewrite the Hamiltonian as, (we are actually looking at the number fluctuations from the background charge density $n_{0}$)
\begin{equation}\label{eq:3}
\mathcal{H_{QR}}=-2J\sum_{\langle ij \rangle }\cos(\theta^{A}_{i}-\theta^{B}_{j})+ U\sum_{i\alpha}(n^{\alpha}_{i})^{2}
,
\end{equation}

For $U/J\gg1$, hopping is suppressed and the system is in the Mott insulating phase \cite{Fisher}. The physics under focus in this work is given in the opposite limit, $J/U\gg1$, where the system is in the super-fluid phase and we shall study the phase fluctuations due to the competing charging energy and the Josephson coupling. For small on-site charging energy $U$, we can write the inequivalent phases as $\theta^{A}_{i}= \theta^{A0}_{i} + \delta \theta^{A}_{i}$ and $\theta^{B}_{j} = \theta^{B0}_{j} + \delta \theta^{B}_{j}$. In the absence of the onsite potential $U$ all the grains should have the same phase $\theta^{A0}_{j}=\theta^{B0}_{j}$ i.e. the grains should be phase coherent. The Hamiltonian is then given by,
\begin{equation}\label{eq:4}
\mathcal{H_{QR}}=-2J\sum_{\langle ij \rangle }\cos(\delta \theta^{A}_{i}-\delta \theta^{B}_{j})+ U\sum_{i\alpha}(n^{\alpha}_{i})^{2}
,
\end{equation}

In this work we will use the Hamiltonian approach. For completeness, we also mention the effective Lagrangian of phase fluctuations can be  formulated as a path integral in the diagonal basis $\theta^{\alpha}_{i}$. The associated action can, then, be written as $\mathcal{S}=\int d\tau \mathcal{L}$ where the Lagrangian is given by \cite{Cha},
\begin{equation}\label{eq:5}
\mathcal{L}= -J\sum_{\langle ij \rangle }\cos(\theta^{A}_{i}-\theta^{B}_{j})+ \frac{1}{4U}\sum_{i\alpha}(\partial_{\tau}\theta^{\alpha}_{i})^{2}
.
\end{equation}

The model Hamiltonian in Eq.~(\ref{eq:4}) is, finally, quadratized by expanding the cosine terms to second order, which is valid for $J/U\gg1$, as the phase fluctuations $\delta \theta^{\alpha}_{i}$ are assumed to be small. We obtain,
\begin{equation}\label{eq:6}
\mathcal{H}'=J\sum_{\langle ij \rangle }[(\delta \theta^{A}_{i}-\delta \theta^{B}_{j})]^{2}+ U\sum_{i\alpha}(n^{\alpha}_{i})^{2}
,
\end{equation}
where we have discarded the constant contribution. We switch to reciprocal space by defining the Fourier transforms of the phase and number operator as $\delta\theta^{\alpha}_{i} = \sum_{{\bf k}}\theta^{\alpha}_{{\bf k}}\exp[i{\bf k}\cdot{\bf R}_{i}]$ and $n^{\alpha}_{i} = \sum_{{\bf k}}n^{\alpha}_{{\bf k}}\exp[i{\bf k}\cdot{\bf R}_{i}]$, giving
\begin{subequations}
\begin{align}
\mathcal{H}'=&
	\sum_{{\bf k}\alpha}
	\biggl\{
		J
		\Bigl(
			3\theta^{\alpha}_{{\bf k}}\theta^{\alpha}_{-{\bf k}}
			-
			\gamma_{\bf k}\theta^{A}_{{\bf k}}\theta^{B}_{-{\bf k}}
			-
			\gamma_{-\bf k}\theta^{A}_{-{\bf k}}\theta^{B}_{{\bf k}}
		\Bigr)
\nonumber\\&
		+
		Un^{\alpha}_{{\bf k}}n^{\alpha}_{-{\bf k}}
	\biggr\} , \label{subeq1}
\\
\gamma_{\bf k}=&
	\sum_{i=1,2,3}e^{i{\bf k}\cdot\boldsymbol{\delta}_i} = 2\cos(\sqrt{3}k_{y}a/2)e^{ik_{x}a/2}+e^{-ik_{x}a}, \label{subeq2}
\end{align}
\end{subequations}

We should note in Eq. (\ref{subeq2}) that $\gamma^{*}_{\bf k}=\gamma_{-\bf k}$. The Hamiltonian in Eq.~(\ref{subeq1}) models two coupled phase oscillations, for which the equation of motion for the two normal modes is,

\begin{equation}\label{eq:8}
\ddot{\phi}^{(1,2)}_{{\bf k}} =-JU(3 \pm |\gamma_{\bf k}|)\phi^{(1,2)}_{{\bf k}},
\end{equation}

where ${\phi}^{(1)}_{{\bf k}}$ and ${\phi}^{(2)}_{{\bf k}}$ are the normal modes of the coupled oscillation in Eq.~(\ref{subeq1}) with $\phi^{(1)}_{{\bf k}}=\frac{(\gamma^{*}_{\bf{k}}\theta^{A}_{\bf{k}}/|\gamma_{\bf{k}}|-\theta^{B}_{\bf{k}})}{\sqrt{2}}$  and $\phi^{(2)}_{{\bf k}}=\frac{(\gamma_{\bf{k}}\theta^{B}_{\bf{k}}/|\gamma_{\bf{k}}|+\theta^{A}_{\bf{k}})}{\sqrt{2}}$. The low energy form of the spectra of these two modes can be found in Sec. II B (See Eq.~ (\ref{eq:13}) and Eq. (\ref{eq:14})), which suggest that ${\phi}^{(1)}_{{\bf k}}$ and ${\phi}^{(2)}_{{\bf k}}$ are massive and massless modes, respectively. By comparing our result with the phase oscillations in two-band superconductors, we identify $\phi^{(1)}_{{\bf k}}$ and $\phi^{(2)}_{{\bf k}}$ as the Leggett mode and Bogoliubov-Anderson-Gorkov (BAG) mode \cite{Shara,Lin}.


\begin{figure}[t]
  \centering
  \includegraphics[width=.38\textwidth,trim= 0 100 0 0,clip ]{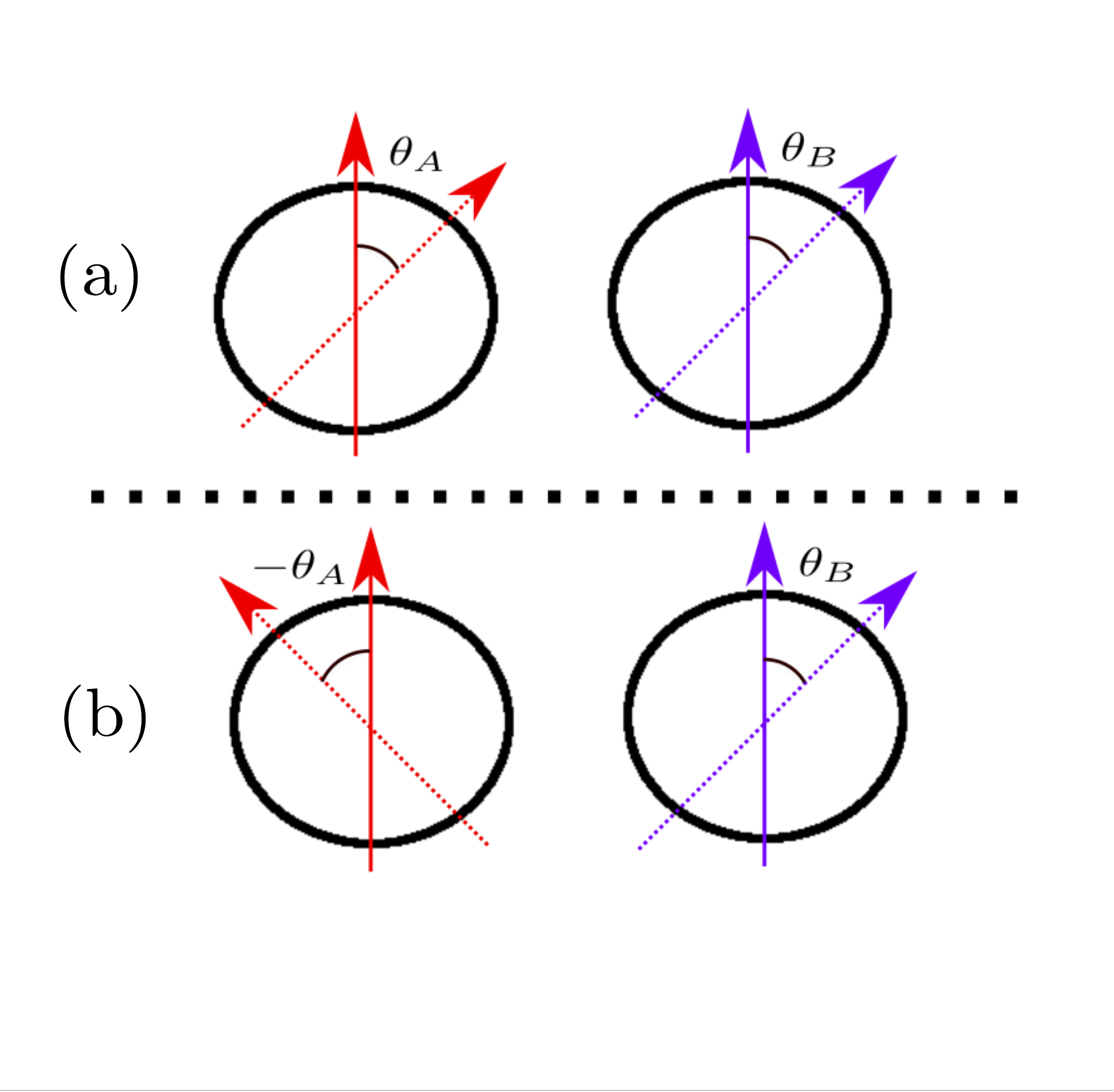}
  \caption{(Color Online) Solid arrows (red and blue) correspond to two inequivalent degenerate phases in the limit $U \sim 0$. (a): Dotted arrows show the collective in-phase (BAG) mode with parallel orientation with finite but small $U$. (b): Dotted arrows show the collective out of phase (Leggett) mode with anti-parallel orientation.}
  \label{fig:Arrow}
\end{figure}

\subsection{Leggett mode and BAG mode}
\begin{figure}[t]
  \centering
  \includegraphics[width=0.99\columnwidth]{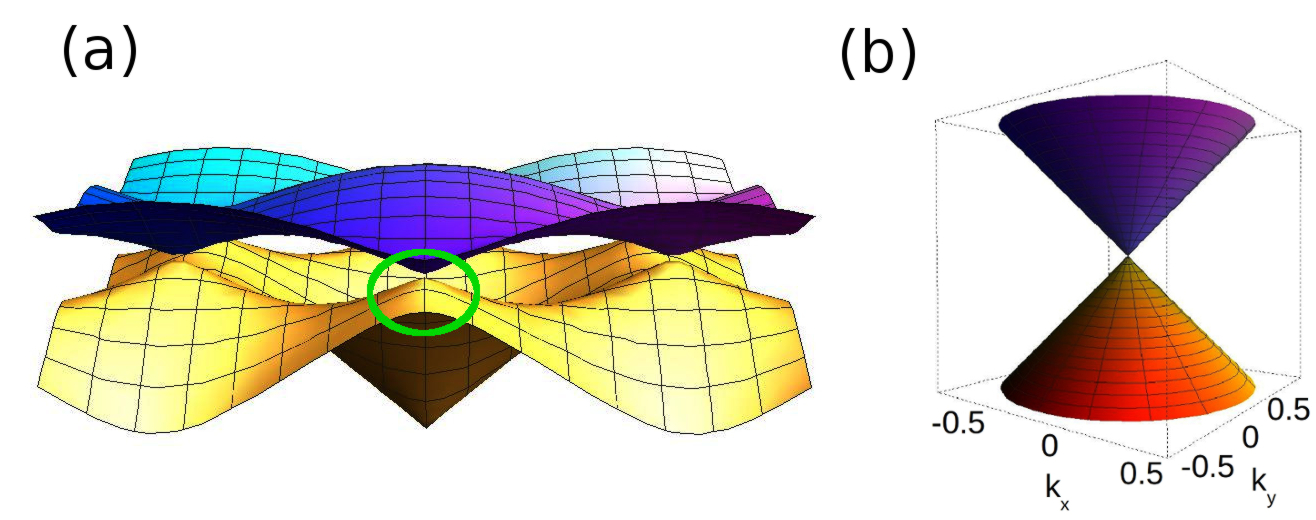}
  \caption{(Color Online) (a) Energy spectra of the bosonic excitations $\omega_{1}({\bf k})$ and $\omega_{2}({\bf k})$ (See Eq. (\ref{eq:9}))  (in units of $\sqrt{JU}$), with J$\approx$ 0.01 eV. and U$\approx$ 0.001 eV (See Section IV). Two modes $\omega_{1}({\bf k})$ and $\omega_{2}({\bf k})$ cross each other at ${\bf K}=2\pi(1,\sqrt{3}/3)/3a$ and ${\bf K'}=2\pi(1,-\sqrt{3}/3)/3a$ in the Brillouin zone and forms a Dirac cone which is shown in green circle. (b) Zoom in of the encircled region in panel (a).}
  \label{fig:Dirac_1}
\end{figure}

Here, we discuss the physical properties of the Leggett and BAG mode in more details. The frequencies of these modes can be obtained from Eq.~(\ref{eq:8}) as,
\begin{equation}\label{eq:9}
\omega_{1,2}({\bf k})=\sqrt{JU(3\pm |\gamma_{\bf k}|)},
\end{equation}
which are plotted in Fig.~\ref{fig:Dirac_1}.

The out of phase, massive, mode $\phi^{(1)}_{{\bf k}}$ with frequency $\omega_{1}({\bf k})$, is identified as the Leggett mode, as indicated in the previous section, while the in phase mode, massless, mode $\phi^{(2)}_{{\bf k}}$ with associated frequency $\omega_{2}({\bf k})$, is associated with the BAG acoustic mode. The existence of two different collective modes is a manifestation of the bipartite lattice structure. A schematic for these modes is shown in  Fig.~\ref{fig:Arrow}, where panel (a) explains graphically the in phase $\phi^{(2)}_{{\bf k}}$ mode and panel (b) explains the out of phase $\phi^{(1)}_{{\bf k}}$ mode. The special feature of the two modes is that they cross each other at the $K$ and $K'$ points, constituting Dirac nodes, see Fig.~\ref{fig:Com1}. The corresponding low energy bosonic excitations following the Leggett and BAG dispersion relations are the main result of this work. We find that it is thus feasible to use an artificial material, made out of superconducting grains, to obtain the bosonic Dirac crossing point for honeycomb lattice. At this point, we should also mention that although we discussed a particular realization for the Dirac physics of Bosons, we claim in general that any other bosonic entity subjected to any bipartite lattice, will exhibit same type of physics.


An important issue is finding out the chiral/helical structure of these collective modes. In most of the conventional Dirac materials the fermionic quasi-particles are chiral. For example, in graphene the massless fermions near the $K$ ($K'$) point are chiral. A relevant quantity to classify the eigenfunctions of the Dirac Hamiltonian (near $K$ point in the Brillouin zone) is the helicity operator. It is defined as the projection of the momentum along the (pseudo)spin direction. The quantum mechanical form of the operator is,
 \begin{equation}\label{eq:10}
\hat{h}=\frac{1}{2}\boldsymbol{\sigma}\cdot\frac{\bf{k}}{|\bf{k}|}
,
 \end{equation}
We will use this definition of the helicity operator to characterise the eigenfunctions of our BDM Hamiltonian. The modes are shown in Fig.~\ref{fig:Dirac_1}. The helicity of the model is found in the low energy regime near the Dirac nodes, shown in Fig. \ref{fig:Com1}. This can be seen easily if we write the free bosonic Hamiltonian in terms of phase variables and diagonalize the phase part,
\begin{align*}
\mathcal{H}=&
	J\sum_{{\bf k}\alpha}
	\biggl(
		3\theta^{\alpha}_{{\bf k}}\theta^{\alpha}_{-{\bf k}}
		-
		\theta^{A}_{{\bf k}}\theta^{B}_{-{\bf k}} \gamma_{\bf k}
		-
		\theta^{A}_{-{\bf k}}\theta^{B}_{{\bf k}} \gamma_{-\bf k}
		+
		Un^{\alpha}_{{\bf k}}n^{\alpha}_{-{\bf k}}
	\biggr)
\\=&
	\sum_{{\bf k}}
	\biggl[
		\epsilon_{1}({\bf k}) \phi^{(1)}_{{\bf k}}\phi^{(1)}_{{\bf -k}}
		+
		\epsilon_{2}({\bf k}) \phi^{(2)}_{{\bf k}}\phi^{(2)}_{{\bf -k}}
		+
		U
		\Bigl(
			n^{A}_{{\bf k}}n^{A}_{-{\bf k}}
			+
			n^{B}_{{\bf k}}n^{B}_{-{\bf k}}
		\Bigr)
	\biggr]
	,
\end{align*}
Here, $\phi^{(1)}_{{\bf k}}$ and $\phi^{(2)}_{{\bf k}}$ are linear combinations of the original $\theta$ variables and $\epsilon_{1,2}({\bf k}) = J(3 \pm |\gamma({\bf k})|)$. In order to find the linear combination we get the unitary matrix for $\theta$ part and consequently write the linear combination as $\phi^{(1)}_{{\bf k}}=\frac{(\gamma^{*}_{\bf{k}}\theta^{A}_{\bf{k}}/|\gamma_{\bf{k}}|-\theta^{B}_{\bf{k}})}{\sqrt{2}}$  and $\phi^{(2)}_{{\bf k}}= \frac{(\gamma_{\bf{k}}\theta^{B}_{\bf{k}}/|\gamma_{\bf{k}}|+\theta^{A}_{\bf{k}})}{\sqrt{2}}$. By introducing new operators $\eta^{(1) \dagger}_{{\bf k}}$ and $\eta^{(2)\dagger}_{{\bf k}}$ (which are linear combinations of $\theta^{(\alpha)}_{{\bf k}}$ and $n^{\alpha}_{{\bf k}}$), we can rewrite the Hamiltonian in Eq. (\ref{subeq1}) as two independent harmonic oscillators according to,
\begin{align}\label{eq:11}
\mathcal{H}=&
	\sum_{{\bf k}}
	\Bigl(
		\omega_{1}({\bf k})\eta^{(1) \dagger}_{{\bf k}}\eta^{(1)}_{{\bf k}}
		+
		\omega_{2}({\bf k})\eta^{(2) \dagger}_{{\bf k}}\eta^{(2)}_{{\bf k}}
	\Bigr),
\end{align}
This is possible since $\phi_{{\bf k}}$ are linear combinations of $\theta_{{\bf k}}$ and since $[\theta^{A}_{{\bf k}}, n^{A}_{-{\bf k}}]=-i$. Using the expansions of $\omega_{1,2}({\bf k})$ discussed in the next subsection, Eqs. (\ref{eq:15}) and (\ref{eq:16}), we write the Hamiltonian near the Dirac point as \cite{Hatsu}
\begin{align}\label{eq:12}
\mathcal{H}_\text{eff}=&
	\omega_0\sigma_0 + v' \boldsymbol{\sigma}\cdot{\bf k}
	,
\end{align}
where $\omega_{0}= \sqrt{3JU}$, $v'= a\sqrt{3JU}/4$, and $\sigma_{0}$ is the $2\times2$ identity matrix. Using the chirality/helicity operator $\hat{h}$ defined in Eq. (\ref{eq:10}), we see that $[\hat{h}, \mathcal{H}_\text{eff}]=0$. Therefore, the eigenfunctions of the Hamiltonian $\mathcal{H}$ in Eq.~(\ref{eq:12}) are also the eigenfunctions of the helicity operator. Hence, we claim that the collective modes are chiral ~\cite{Wehling}.

\begin{table}[t]
  \centering
  \caption[font=large]{\bf $\gamma_{\bf k}$ near $K$ and $\Gamma$ point}
  \begin{tabular}{ccc}
    \hline \hline
    \textbf{Function} & \textbf{$\Gamma$ Point} & \textbf{$K$ Point}\\
    \hline
    $\gamma_{\bf k}$ & $\left(3-\frac{3}{4}a^{2}|\bf{k}|^2\right)$  & $\frac{3a}{2}(k_x+ik_y)e^{i\frac{5\pi}{6}}$\\
    \hline
    $|\gamma_{\bf k}|$ & $\left(3-\frac{3}{4}a^{2}|\bf{k}|^2\right)$  & $\frac{3a}{2}|\bf{k}|$\\
    \hline
  \end{tabular}
\end{table}

\subsection{Low energy behavior of the excitations}
In this section we investigate the low energy theory of the bosonic excitations near the $\Gamma$ and $K$ points in the Brillouin zone. It is important to distinguish the Leggett mode from BAG mode by looking at the forms of the spectra near the $\Gamma$ point. The acoustic mode in Eq. (\ref{eq:13}) is identified as the BAG mode ({\it i.e.} $\omega_{2}({\bf k})$ is the BAG mode ) and the Leggett mode is $\omega_{1}(\bf{k})$ (by examining Eq. (\ref{eq:14})) ~\cite{Shara}.

\subsubsection{BAG mode near $\Gamma$ point}
For ${\bf k}={\bf q}$ where $|{\bf q}|\ll1/a$ we have, see Table~II,
\begin{equation}\label{eq:13}
 \omega_{2}^{2}({\bf q}) \simeq \frac{3JU}{4}\,a^{2}|{\bf q}|^{2}+\mathcal O({\bf q}^{4})
 ,
\end{equation}
The linear low energy dispersion relation suggests that the BAG mode is acoustic and that this mode has a group velocity around $\Gamma$ point with $v_g \sim a\sqrt{3JU}/2$. $v_g$ is defined as the norm of $|\nabla \omega({\bm k})|$.

\begin{figure}[t]
  \centering
  \includegraphics[height=1.6 in ,width=3.5 in]{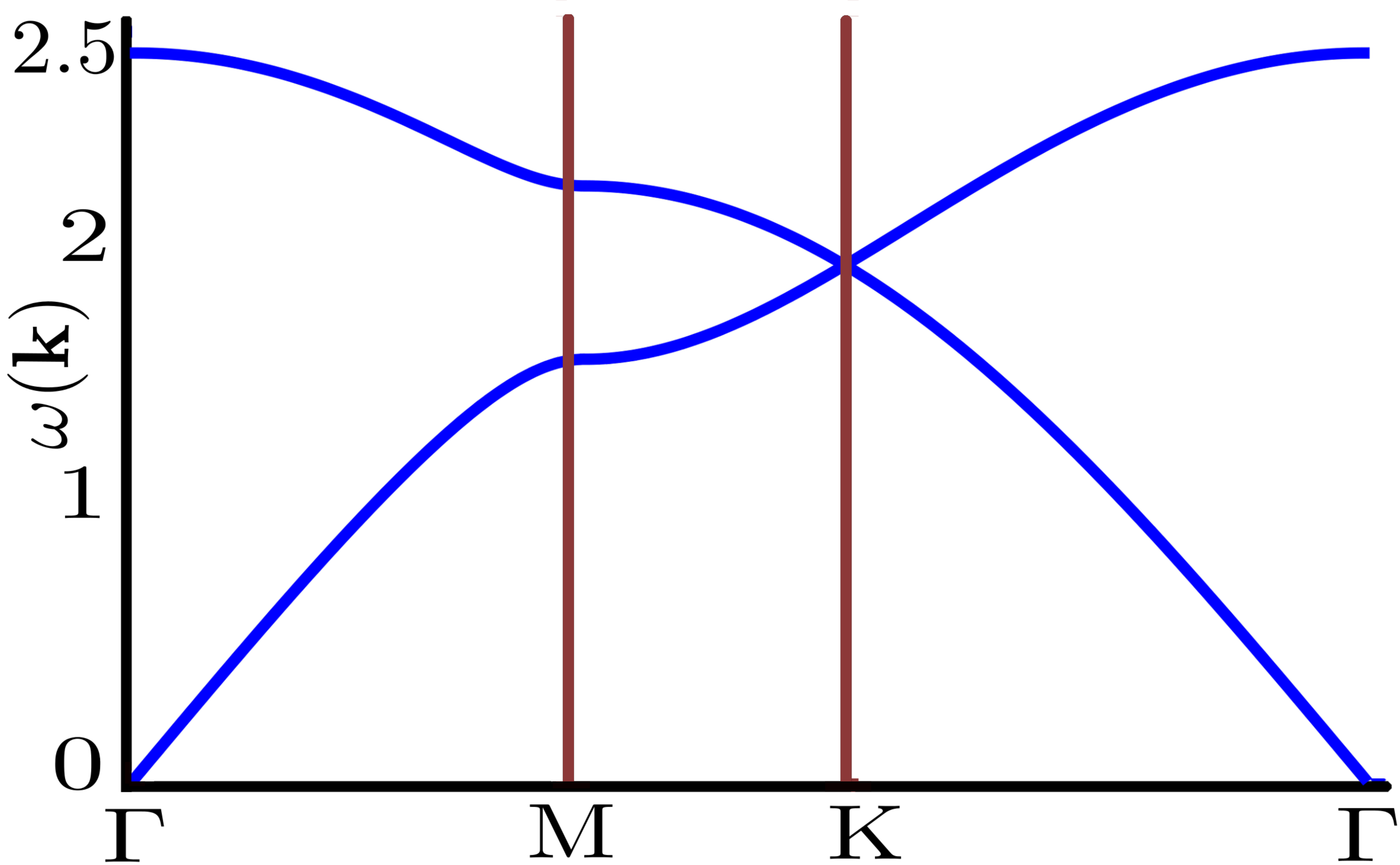}
  \caption{The spectra for bosonic collective modes (in units of $\sqrt{JU}$) $\omega_{1}({\bf k})$ (Leggett, top band)  and $\omega_{2}({\bf k})$ (BAG, down band)  as traversed from high symmetry points $\Gamma $ to $M$ to $K$ to $\Gamma $. The spectra touch each other at $K$ and $K'$ points in the Brillouin zone and form Dirac cones.}
  \label{fig:Com1}
\end{figure}

\subsubsection{Leggett mode near $\Gamma$ point}
The dispersion relation of the Leggett mode near the $\Gamma $ point is massive and with negative curvature. For ${\bf k}={\bf q}$ where $|{\bf q}|\ll1/a$ we have, see Table.~II,
\begin{equation}\label{eq:14}
  \omega_{1}^{2}({\bf q}) \simeq JU\biggl(6-\frac{3a^{2}}{4}|{\bf q}|^{2}\biggr)+\mathcal O({\bf q}^{4})
  ,
\end{equation}
suggesting that the Leggett mode is an optical mode. It is clear from the dispersion relation (Eq. (\ref{eq:14})) that this mode has a negative curvature near the $\Gamma$ point and located at high energy compared to the BAG mode.

\subsubsection{BAG mode near K and $K'$ point}
In contrast to the Leggett mode low energy spectrum, the BAG mode gives the following linear dispersion near ${\bf k}={\bf K}+{\bf q}$ where $|{\bf q}|\ll1/a$ as, see Table~II,
\begin{equation}\label{eq:15}
 \omega_{2}({\bf q})\simeq
 	\sqrt{3JU}
	\biggl(
		1-\frac{a|{\bf q}|}{4}
	\biggr)+\mathcal  O({\bf q}^2)
	,
\end{equation}
The group velocity of this mode is $v_g \sim a\sqrt{3JU}/4$ which also exhibits energy-shifted Dirac point compared to graphene dispersion ~\cite{Wehling,Castro}.

\subsubsection{Leggett mode near K and $K'$ point}
The dispersion relation of the Leggett mode near the Dirac point ${\bf K}=2\pi(1,\sqrt{3}/3)/3a$ in Brilloiun zone for ${\bf k}={\bf K}+{\bf q}$ where $|{\bf q}|\ll1/a$, see Table~II, is given by
\begin{equation}\label{eq:16}
  \omega_{1}({\bf q})\simeq
  	\sqrt{3JU}
	\biggl(
		1+\frac{a|{\bf q}|}{4}
	\biggr)
	+\mathcal O({\bf q}^2)
	,
\end{equation}
The Dirac point is shifted in energy by a term proportional to $\sqrt{JU}$ and the group velocity $v_g \sim a\sqrt{3JU}/4 $.  We notice that the charging energy shifts the position of the Dirac point in energy space. Note that energy shift is same for BAG and Leggett modes and thus their spectra touch  at the K and $K'$ points, forming the Dirac cone. We also note that both BAG and Leggett modes have same group velocity only differing in sign.

\subsubsection{Gap opening at $K$ and $K'$ point}

\begin{figure}[t]
\begin{center}
\includegraphics[width=.99\columnwidth]{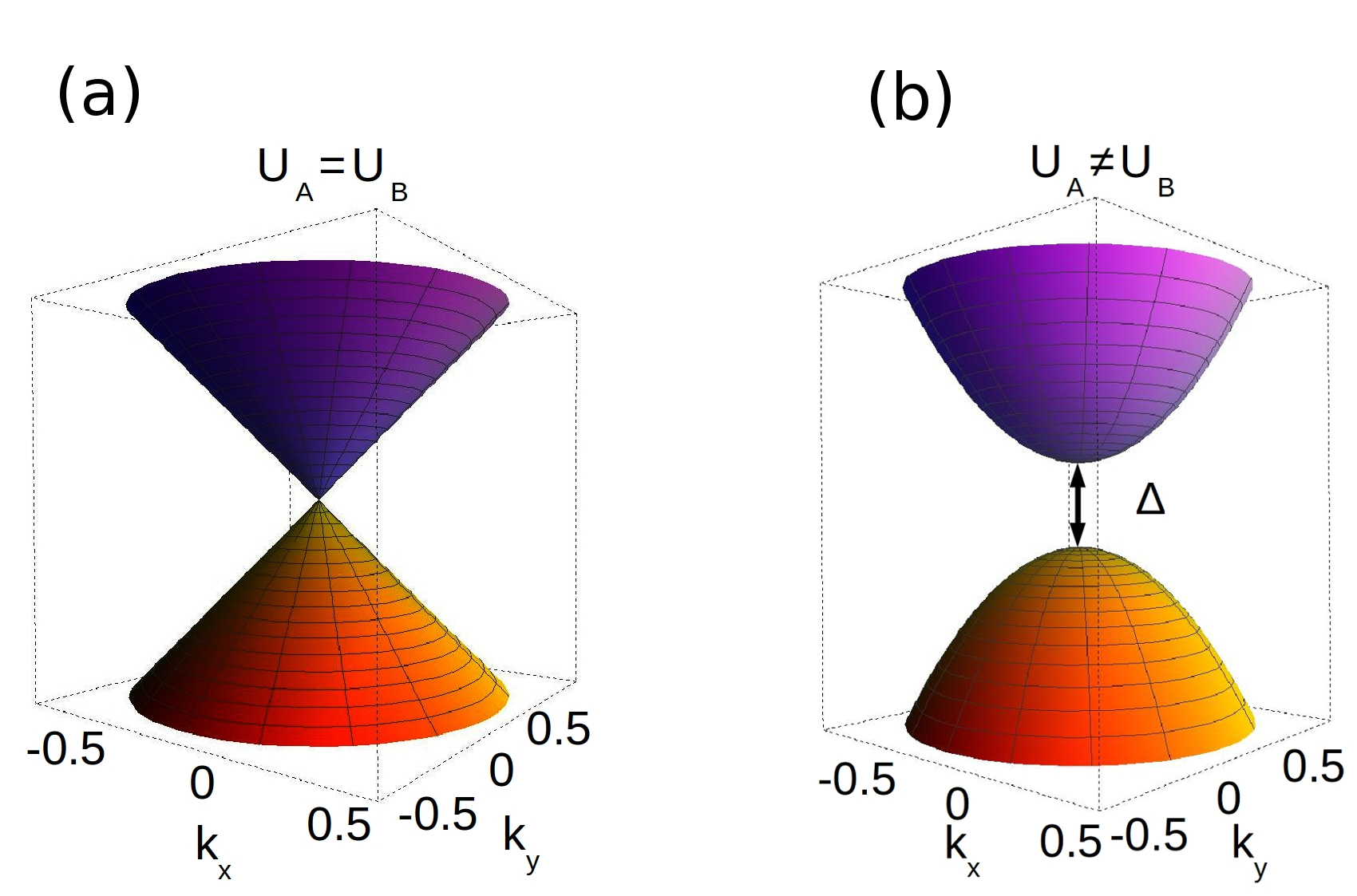}
\end{center}
\caption{(Color online) (a) The linear dispersion relation near  for same on-site energy $U_{A} = U_{B}$. (b) Different on-site energy $U_{A} \neq U_{B}$ leads to opening of gap ($\Delta$) near the $K$ point in the Brillouin zone for the bosonic modes.}
\label{fig:Gap}
\end{figure}

For different on-site energies $U_{A} \neq U_{B}$, we obtain the following dispersion relation for the BAG and Leggett mode near the Dirac point $K$,

\begin{subequations}
\begin{align}
 \omega_{1}^{2}({\bf q}) & \simeq 3JU_{A}+\frac{3JU_{A}U_{B}}{2(U_{A}-U_{B})}a^{2}|{\bf q}|^{2}+\mathcal O({\bf q}^2) \label{subeq3},
 \\
 \omega_{2}^{2}({\bf q}) & \simeq 3JU_{B}-\frac{3JU_{A}U_{B}}{2(U_{A}-U_{B})}a^{2}|{\bf q}|^{2}+\mathcal O({\bf q}^2)\label{subeq4}
 ,
\end{align}
\end{subequations}
The gap develops at the Dirac point since (i) $U_A\neq U_B$ and (ii) the modes are shifted in energy by $\sqrt{3JU_{A}}$ and $\sqrt{3JU_{B}}$ respectively (see Fig.~\ref{fig:Gap} and Eq. (\ref{eq:15}) and Eq. (\ref{eq:16})).

In graphene \cite{Castro} the spectrum near the Dirac point is linear and a gap can be opened by breaking the sublattice symmetry. Analogously, in the bosonic Dirac spectrum a gap is opened whenever the on-site charging energy for the two sublattices $U_{A}$ and $U_{B}$ are distinct. Tunability of the bosonic spectrum due to the gap opening allows using these materials for thermal, optical and transport applications. This feature is also applicable for other realizations of Bosonic Dirac materials by breaking the sublattice symmetry. \cite{Wehling,Vafek}

\begin{figure}[ht]
\centering
\subfigure[(Color Online) The evolution of the collectives modes ($\omega({\bm k})$ in unit of $\sqrt{JU}$) as we turn on the onsite potential interaction $U$. Without any interaction the lowest band near $\Gamma$ point is quadratic and becomes linear as $U$ changes from 0. Dirac point moves up in energy as one increases $U$.]{
\includegraphics[width=0.95\columnwidth]{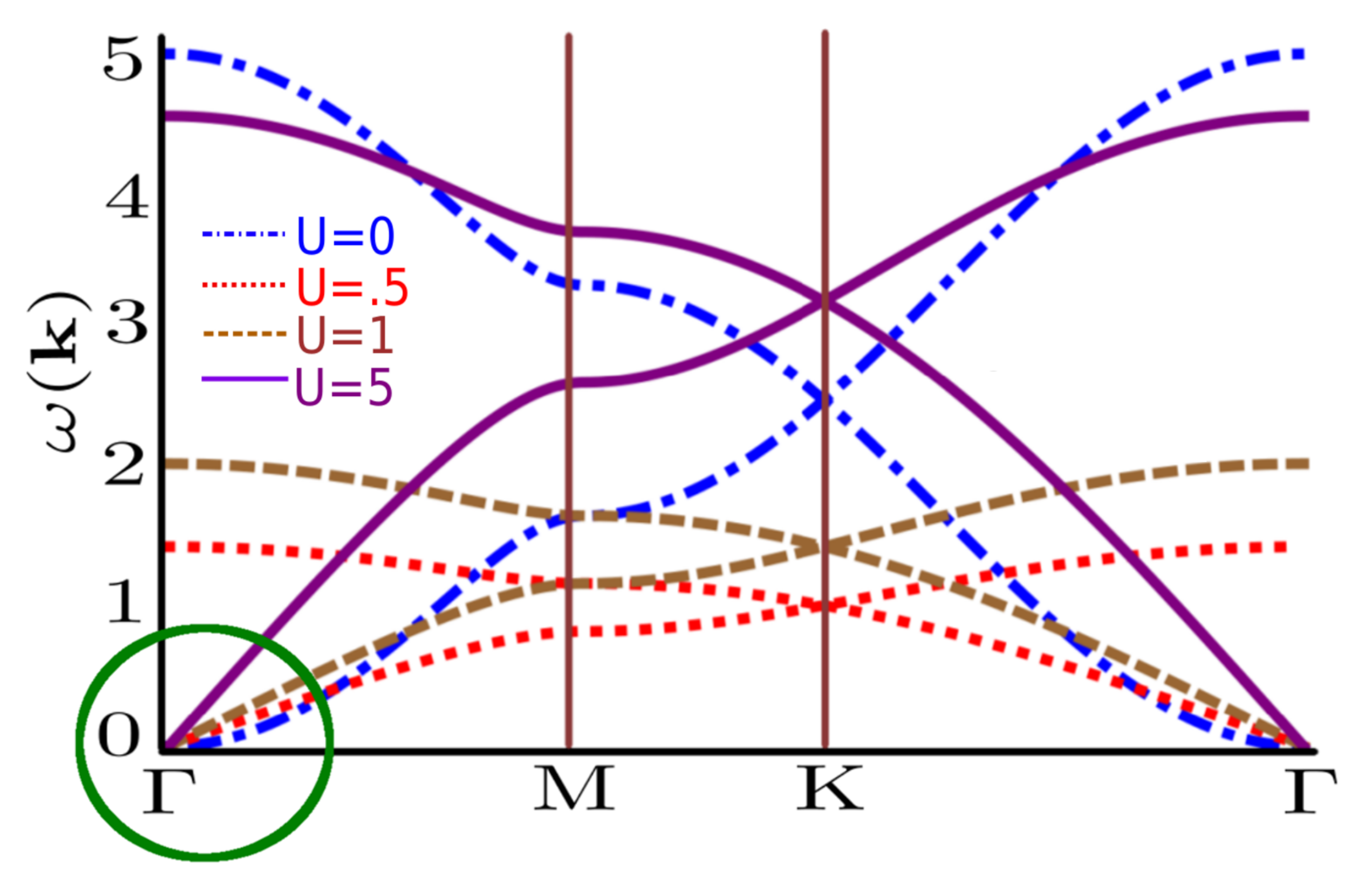}
\label{fig:HS}
}\\
\subfigure[(Color Online) The zoom in of the region encircled in green, in Fig.\ref{fig:HS} near $\Gamma$.]{
\includegraphics[width=0.95\columnwidth]{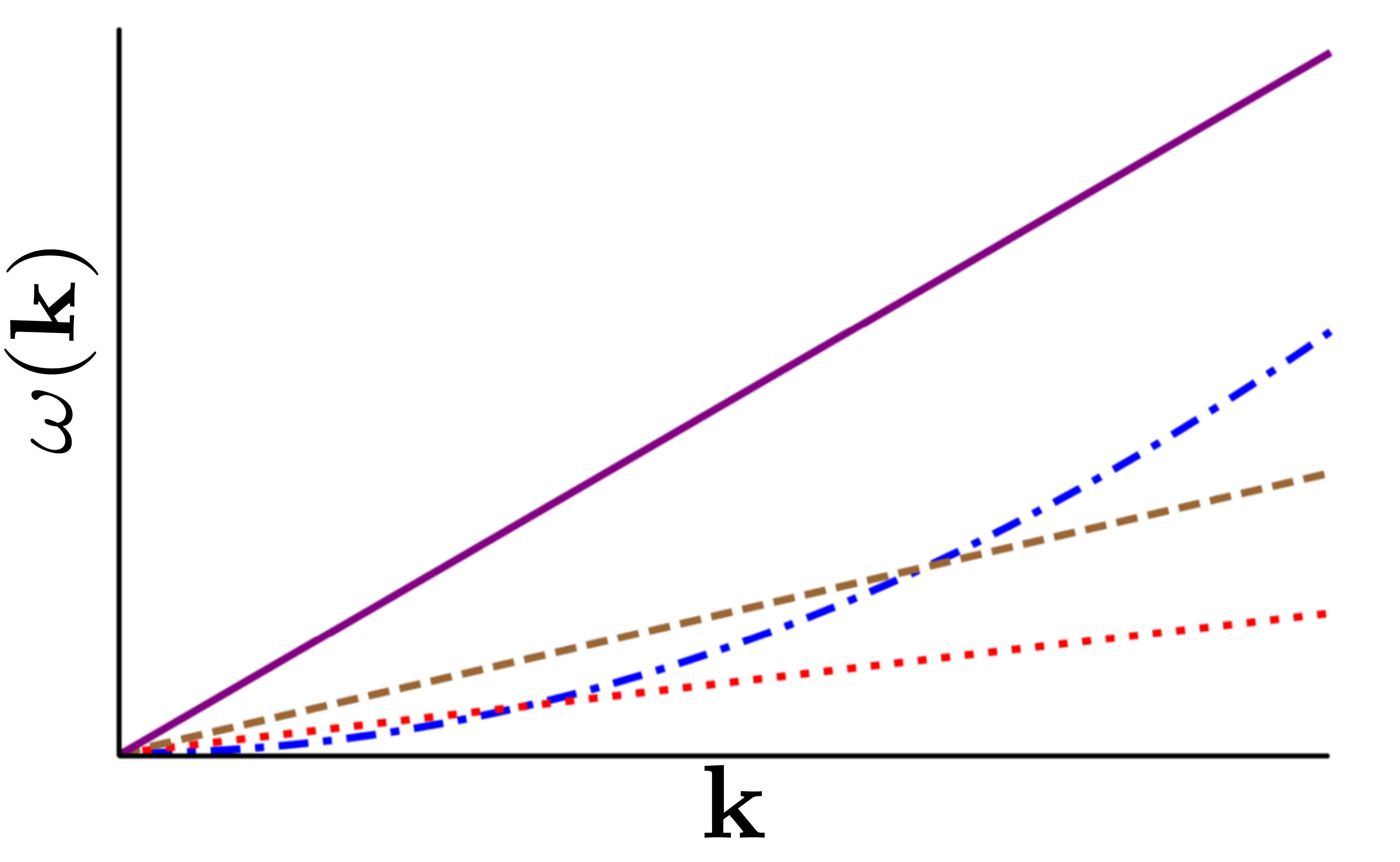}
\label{fig:HSS}
}
\caption[Optional caption for list of figures]{An illustration of the evolution of the bands (Leggett and BAG) as one changes $U$ from 0 to a finite value.}
\label{fig:subfigureExample1}
\end{figure}

For completeness we also present the evolution of bosonic modes as one changes $U$ in Fig. \ref{fig:HS}. The dispersion relation for the lowest band near the $\Gamma$ point is shown in Fig. \ref{fig:HSS}. For $U=0$ the lowest band is quadratic near the $\Gamma$ point. As one turns on the interaction $U$ , the BAG mode becomes linear or acoustic. In our approximate solution where we focus on phase fluctuations only, we observe initial linear slope, as fully expected for weakly interacting bose gas. We recall that the dispersion relation for weakly interacting bose gas \cite{Pethick} has a leading order dispersion, linear in momentum with a coefficient depending on the interaction strength and sub-leading term, cubic in momentum, whose coefficient is independent of the interaction strength. The apparent discontinuity between U=0 and finite U behaviour of BAG mode is a manifestation of $\mathcal O(2)$ expansion in momentum k of dispersion relation for the particular phase approximation in our model which is very similar to the approximation in Bogoliubov perturbation theory.

\section{Microscopic Hamiltonian for the granular superconductor: Inter-grain Interaction}

In the previous section we discussed Dirac-like bosonic collective oscillations for phase fluctuations of the 2D superconducting grains with on-site charging energy $U$. Here, we shall take into account the inter-granular interactions. Practically, the grains are assumed to be large and contain a large number of charged Cooper pairs. Therefore, one would expect an inter-grain long-range Coulomb interaction. As we shall see in Appendix B, the inclusion of the Coulomb interaction qualitatively changes the properties of the two collective modes. The acoustic BAG mode becomes gapped as can be seen by comparing the low energy expansion of the dispersion relation in Eq.~(\ref{eq:13}) (without inter-grain Coulomb interaction) and Eq.~(\ref{eq:33}) in Appendix B (with inter-grain Coulomb interaction) and also comparing Fig.~\ref{fig:Com1} and Fig.~\ref{fig:Com_1}. The Leggett mode remains qualitatively the same  but the parameters are changed which can be seen  by comparing Eq.~(\ref{eq:14}) (without inter-grain interaction) and Eq.~(\ref{eq:22}) (with inter-grain interaction) and also comparing the Fig.~\ref{fig:Com1} and Fig.~\ref{fig:Com_1}. The model is analytically intractable if we consider the complete long range Coulomb interaction $\sum_{ ij \alpha \beta}  U'_{ij}n^{\alpha}_{i} n^{\beta}_{j}$ where $ij$ summation extends over the whole lattice sites ($\alpha ,\beta$ summation extends over the sub-lattice indices). Therefore, in this section we will consider nearest neighbour local interaction $U'$ and will find out its effect on the behavior of the collective modes. The extended quantum rotor model for the inter-grain interaction in the system is,
\begin{align}\label{eq:18}
\mathcal{H}&=-2J\sum_{\langle ij \rangle }[cos(\theta^{A}_{i}-\theta^{B}_{j})]+ U\sum_{i\alpha}(n^{\alpha}_{i})^{2} + U'\sum_{ \langle ij \rangle \alpha \beta } n^{\alpha}_{i} n^{\beta}_{j}
,
\end{align}

As discussed in the beginning of this section, we give an approximate description of the effect of the long range Coulomb interaction $U'_{ij}$ in Appendix B. Assuming that both $U$ and $U'$ are smaller than $J$, we linearise the above Hamiltonian in Eq. (\ref{eq:18}) by expanding the cosine term to quadratic order, such that this model becomes analytically solvable. Fourier transforming the phase and number variables, we obtain the effective Hamiltonian of the phase fluctuations as,
\begin{align}\label{eq:19}
\mathcal{H}=
	\mathcal{H}'+ \frac{U'}{2}\sum_{{\bf k}\alpha} \left( \gamma_{\bf k}n^{A}_{{\bf k}} n^{B}_{-{\bf k}}+ \gamma_{-{\bf k}}n^{A}_{-{\bf k}} n^{B}_{{\bf k}} \right)
,
\end{align}
where $\mathcal{H}'$ is the same as in Eq.~(\ref{subeq1}). Inclusion of this local interaction between the grains in the model leads to a few qualitative changes in the behavior of the two collective modes (we call them modified Leggett and modified BAG ) which can be seen by carefully examining the Eqs. (\ref{eq:13})-(\ref{eq:16}) and Eqs. (\ref{eq:21})-(\ref{eq:24}). It is important to see that the modified BAG mode is still an acoustic mode (Eq. (\ref{eq:21})).

%
%

\begin{figure}[t]
  \centering
  \includegraphics[width=1\columnwidth]{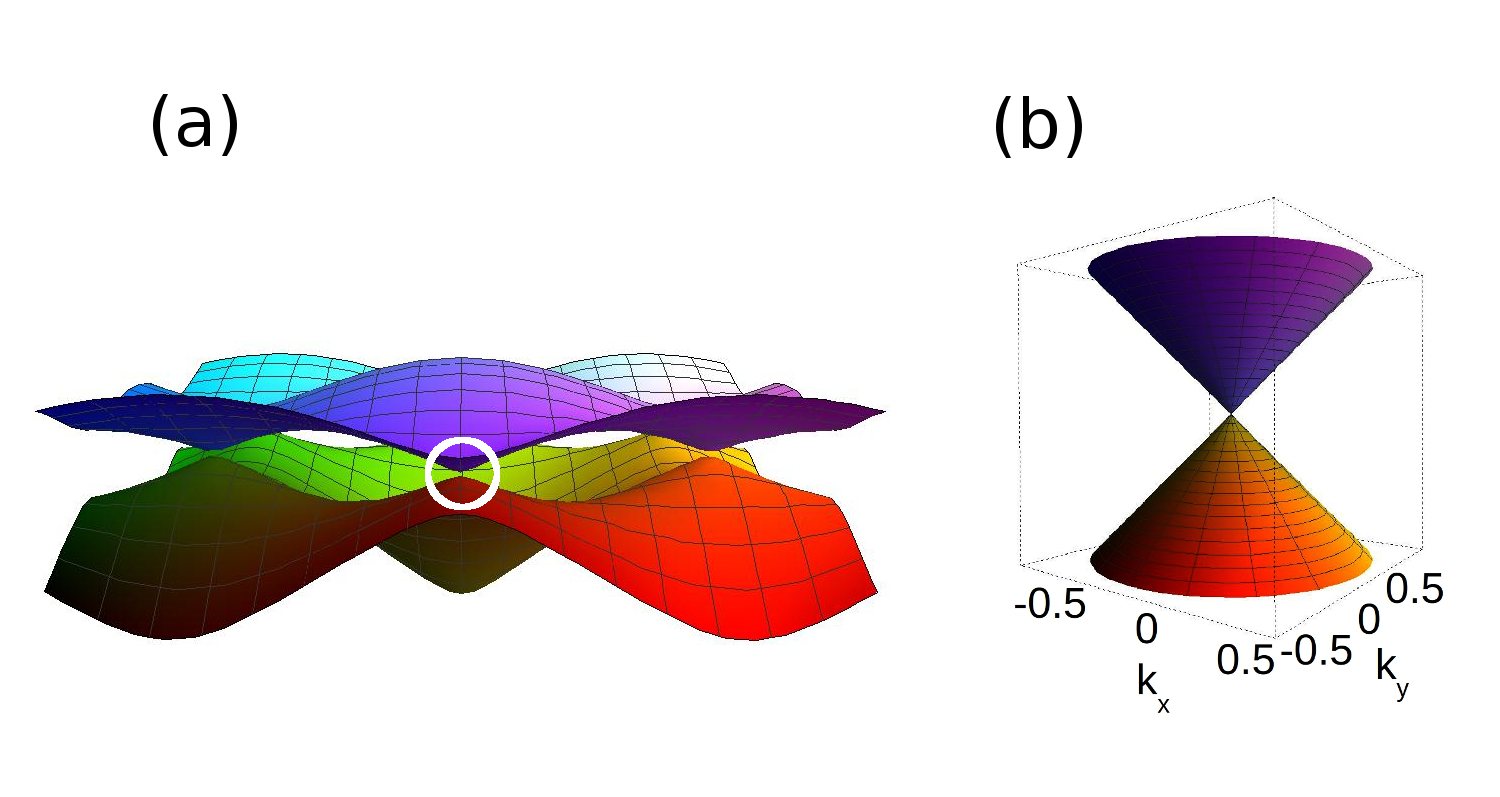}
  \caption{(Color Online) (a) Energy spectra of the bosonic excitations $\omega_{1}({\bf k})$ and $\omega_{2}({\bf k})$ (See Eq. (\ref{eq:20}), in units of $\sqrt{JU}$), with J$\approx$ 0.01 eV, $U/J$ $\sim$ 0.1 and  $U' \sim 8\cdot10^{-4}$ eV (Sec. IV). Two modes $\omega_{1}({\bf k})$ and $\omega_{2}({\bf k})$ cross each other at ${\bf K}=2\pi(1,\sqrt{3}/3)/3a$ and ${\bf K'}=2\pi(1,-\sqrt{3}/3)/3a$ points in the Brillouin zone and forms a Dirac cone which is shown in white circle.  (b) Zoom in of the encircled region in panel (a).}
  \label{fig:Dirac_4}
\end{figure}

%
%
 \begin{figure}[b]
  \centering
  \includegraphics[height=1.6 in ,width=3.5 in]{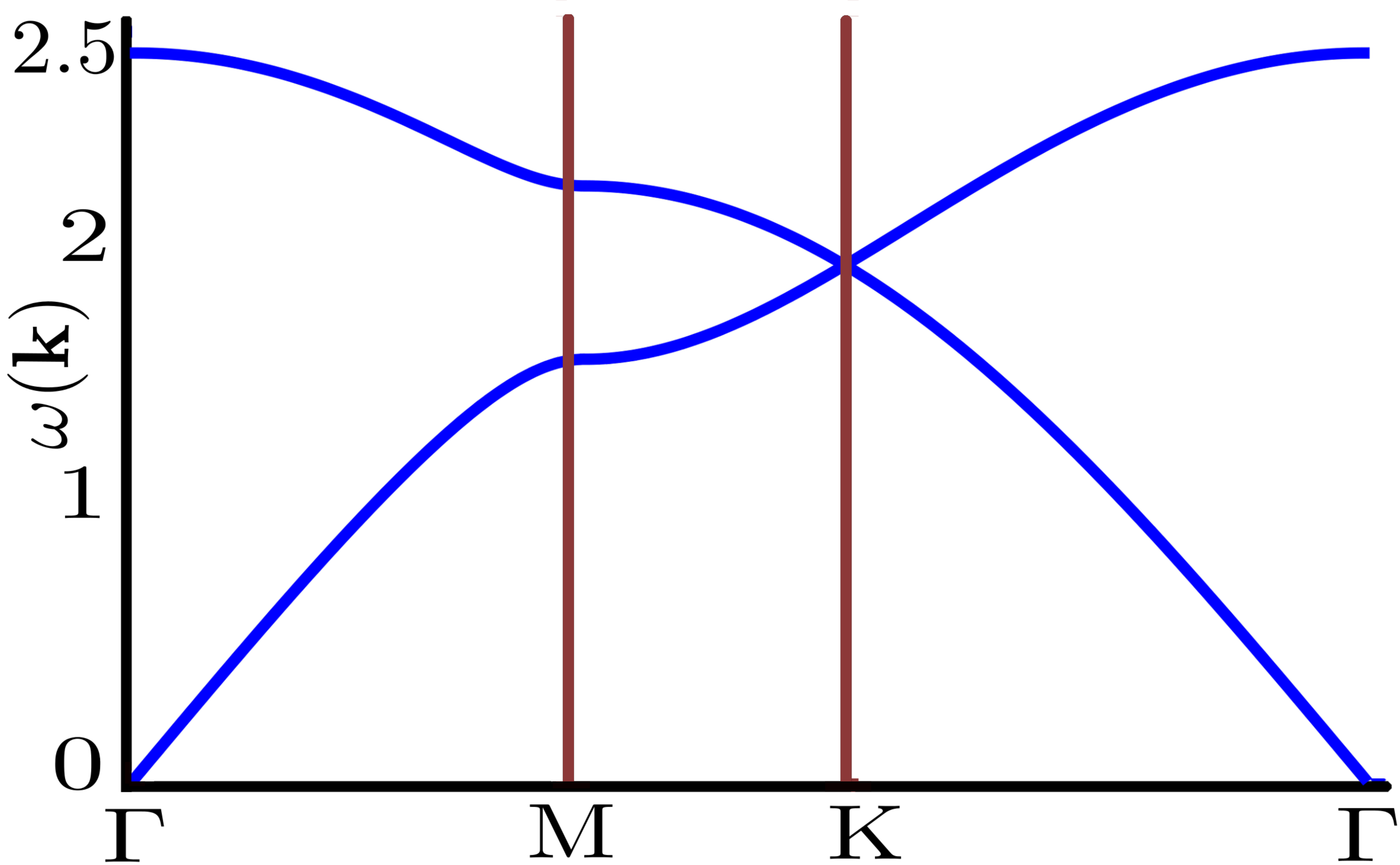}
  \caption{(Color Online) The spectra for bosonic collective modes (in units of $\sqrt{JU}$)  $\omega_{1}({\bf k})$ (modified Leggett, top band) and $\omega_{2}({\bf k})$ (modified BAG, down band)  as traversed from high symmetry points $\Gamma $ to $M$ to $K$ to $\Gamma $. The interaction $U'$ is finite. The modes $\omega_{1}({\bf k})$ and $\omega_{2}({\bf k})$ cross each other at $K$ and $K'$ points in the Brillouin zone and form Dirac cones.}
  \label{fig:Com2}
 \end{figure}

We find that the two collective phase modes in this case also cross each other at the Dirac points ${\bf K}$, ${\bf K}'$, (Figs.~\ref{fig:Dirac_4} and \ref{fig:Com2}) and, as discussed in the previous Sec. II A, these modes are chiral following the same argument. The frequencies of these two modes are calculated from the Hamiltonian Eq.~(\ref{eq:19}) and we obtain,
\begin{equation}\label{eq:20}
\omega_{1,2}^{2}({\bf k})=J\left(3U \mp \frac{3U'}{2}|\gamma_{\bf k}|\pm U|\gamma_{\bf k}|-\frac{U'}{2}|\gamma_{\bf k}|^2\right)
,
\end{equation}
For $U'=0$ the frequencies $\omega_{1,2}({\bf k})$ reduce to those of Eq.~(\ref{eq:9}). The mode crossing is shown explicitly in Fig.~6 in the presence of neighbouring grain interaction $U'$. The crossing of two bands along the high symmetry points in the Brillouin zone is shown in Fig.~\ref{fig:Com2}. In the next section, we focus on the dispersion relations of these two modes near the Dirac point and extract the Dirac physics.

\subsection{Low energy behavior of the excitations}
In this section we describe the low energy behavior of the modified bosonic excitations near $\Gamma$ and $K$ points in the presence of the nearest neighbour interaction $U'$. Therefore, we do not see any considerable quantitative changes except some renormalization of the parameters. In Appendix B we give an intuitive argument for the modification of the behavior of the acoustic BAG mode near $\Gamma$ point. We actually identify the modes by looking at their forms near the $\Gamma$ point.

\subsubsection{Modified BAG mode near $\Gamma$ point}
The dispersion relation of the previously discussed BAG mode near $\Gamma$ point in Brilloiun zone, for ${\bf k}={\bf q}$ where $|{\bf q}|\ll1/a$, see Table~II and Eq.~(\ref{eq:20}), is given by
\begin{equation}\label{eq:21}
 \omega_{2}^{2}({\bf q})\simeq
 	\frac{3JU}{4}
	\left(1+\frac{3U'}{U}\right) a^{2}|{\bf q}|^{2} +\mathcal{O}(|{\bf q}|^{4})	
,
\end{equation}
We see that when there is no interaction the frequency matches Eq. (\ref{eq:13}). We also see a change in the group velocity $v_{g} \sim a\sqrt{3JU(1+3U'/U)}/2$ compared to the BAG mode without the interaction $U'$.

\subsubsection{Modified Leggett mode near $\Gamma$ point}
The dispersion relation of the modified Leggett mode near the $\Gamma$ point, for ${\bf k}={\bf q}$ where $|{\bf q}|\ll1/a$, see Table~II and Eq.~(\ref{eq:20}), is given by
\begin{align}\label{eq:22}
 \omega_{1}^{2}({\bf q})\simeq&
 	6JU - 9JU'- \frac{3JU}{4}\bigg(1-\frac{9U'}{2U} \bigg) a^{2}|{\bf q}|^{2} +\mathcal{O}(|{\bf q}|^{4})
,
\end{align}
like the Leggett mode in the previous section, this mode is have a negative curvature and located at high energy compared to modified BAG mode. For $U'=0$, the dispersion reduces to the Leggett mode frequency in Eq.~(\ref{eq:14}). While the inter-granular interaction renormalizes the parameters in the Leggett mode, the qualitative properties remain essentially unaffected (See Eq.(\ref{eq:14})).

\subsubsection{Modified BAG mode near K and $K'$ point}
The dispersion relations of the previously discussed BAG modes near the point ${\bf K}$ in the Brillouin zone for ${\bf k}={\bf K}+{\bf q}$ where $|{\bf q}|\ll1/a$, see Table~II and Eq.~(\ref{eq:20}), is given by
\begin{align}\label{eq:23}
\omega_{2}({\bf q})\simeq
 	\sqrt{3JU}
	\biggl[
		1
		-
		\biggl(\frac{1}{4}-\frac{3U'}{8U}\biggr)a|{\bf q}|
	\biggr]
	+
	\mathcal O({\bf q}^2)
,
\end{align}
We see that the modified BAG mode energy is shifted exactly in the same manner as in Eq. (\ref{eq:15}). The group velocity is given by $v_g \sim a\sqrt{3JU}(1-3U'/2U)/4$.

\subsubsection{Modified Leggett mode near K and $K'$ point}
The dispersion relations of the Leggett mode near the point $K$ in the Brillouin zone for ${\bf k}={\bf K}+{\bf q}$ where $|{\bf q}|\ll1/a$, see Table~II and Eq.~(\ref{eq:20}), is given by
\begin{align}\label{eq:24}
\omega_{1}({\bf q})\simeq
 	\sqrt{3JU}
	\biggl[
		1
		+
		\biggl(\frac{1}{4}-\frac{3U'}{8U}\biggr)a|{\bf q}|
	\biggr]
	+
	\mathcal O({\bf q}^2)
,
\end{align}
The modified Leggett mode is also shifted in energy by a term depending on the charging energy. It is important to note that energies of both the modified BAG and modified Leggett modes are shifted by the same amount and hence they touch each other at $K$ and $K'$ points in the Brillouin zone. This mode has a group velocity of $v_g \sim  a\sqrt{3JU}(1-3U'/2U)/4$. We see that modified BAG and Leggett modes have the same group velocity $v_g$ only differing in sign, as in Sec. II.

\section{Parameter choice for the granular model and role of disorder}
We described the Dirac nodes in the context of bosonic excitations. We now turn to the range of parameters that can be tuned in the class of granular superconductors.  We typically assumed $J=0.01$ eV and $U/J=1/10$. For these parameters we will get the typical velocity of boson modes near $K,K'$ to be, (for $U'=0$)
 \begin{align}\label{eq:25}
  v_g \sim 5\,meV\,a,
 \end{align}
where $a$ is unit cell size, which we expect to be on the range of microns.  By tuning the inter-grain distance and also the granular size we can have some range of choice of the parameter for $J$, $U$, and $U'$. This facilitates changes in the group velocity of the modes near the Dirac points. Moreover, the gap can be opened at the Dirac points $K$ and $K'$ when the on-site charging energies in the $A$- and $B$-sublattices are chosen differently as $U_{A} \neq U_{B}$, see  Fig.~\ref{fig:Gap}. If we assume $U_{A} \sim 3\:meV$ and $U_{B} \sim 1\:meV$, the gap magnitude will be on the order of (for $J \sim 0.01$ eV),
 \begin{align}\label{eq:26}
 \Delta \sim \sqrt{3JU_{A}}-\sqrt{3JU_{B}} \sim 4 meV,
  \end{align}
 which would make it easily observable in spectroscopies. Optical absorption, local tunnelling  probes and transport will be sensitive to the gap opening at the Dirac nodes and can thus provide experimental evidence for the Dirac nature of the bosonic modes.

We should also mention effects of lattice disorder. As we are analyzing the artificial lattice that can not be prepared perfectly, we point out that lattice disorder will lead to on-site potential variations and inter-grain coupling energy fluctuations. All these effects will lead to the modification of the bosonic spectrum. One can separate the effect of disorder in two categories. On one hand the \emph{on-site} disorder will lead to localized bosonic excitations, as would be the case for the fermionic analogue \cite{Wehling}, where local perturbations of the on-site potential will induce local single boson resonances. On the other hand, the inter-grain potential energy variations will induce changes in the gaps at $K,K'$ points and smear them. Both of these effects are important and would need to be addressed in detail. The analysis of the role of disorder is a subject of a separate investigation and is deferred for a separate publication.

\section{Discussion and Conclusion}
We presented the case for bosonic excitations on the honeycomb lattice that lead to the Dirac node in bosonic dispersion. To carry out specific calculations and illustrate the formation of the Dirac node we used the specific example of superconducting grains forming honeycomb lattice.  In this work we have proposed to use granular 2D superconductors as a platform to realize bosonic Dirac materials (BDM). While the calculations are specific to the case of Josephson network we argue that there are  universal statements one can make. We would like to point out that universal statement about having a Dirac cone in such two dimensional materials can be made with the help of Von Neumann-Wigner theorem~\cite{WangJ}. We  To this end we have solved the real particle Bose-Hubbard model in a 2D film of superconducting grains, arranged in honeycomb lattice. We find that in the superfluid phase we have a two component superfluid with  collective phase oscillations that exhibit Dirac points in the spectrum. In contrast to graphene \cite{Castro} and other known Dirac materials \cite{Wehling,Vafek}, these Dirac modes are bosonic excitations.  These modes represent the Bogoliubov-Anderson-Gorkov (BAG) mode and Leggett modes that touch at the $K,K'$ points of the Brillouin zone. We also find that the bosonic modes are chiral in similarity with the conventional fermionic Dirac materials. The proposed realization of the BDM also opens up a route to design multicomponent superconducting materials using the bipartite nature of honeycomb lattices. The two sublattice  structure of the granular superconductor supports two component superconducting state, where we focused only on its phase dynamics here.

Another interesting observation is that different local interactions $U_A$ and $U_B$ opens the gap in the Dirac spectrum and thus allows one to control the bosonic excitation spectrum. Extra advantage in the case of artificial SC grains is the tunability of the grain sizes and spacings that will lead to tunable spectra and hence make the determination of the Dirac nature of the spectra easier to accomplish. To access the Dirac point and probe the properties at the energies on that scale one would need to perform inelastic scattering measurements like inelastic neutron and sound scattering among other probes.

We left outside of this work important questions that would be needed to be addressed in a due course. Role of phase fluctuations and vortex excitations, BCS-BEC crossover \cite{Zhao} BKT transition and effects of the charge ordering at commensurate fillings would be interesting directions to pursue. As we already indicated, another important topic would be to investigate the role of disorder. We plan to address these questions in subsequent work.

\section{Acknowledgement}
We are grateful to J. Lidmar and M. Wallin  and  for many important discussions regarding the idea of the works. This work was supported by US DOE BES E304. H.{\AA}. acknowledge the Knut and Alice Wallenberg foundation for financial support (Grant No. KAW-2013.0020).  Work at KTH and Uppsala was supported by ERC DM 321031 and the Swedish Research Council (Vetenskapsr{\aa}det).

\appendix
\section{The quantum rotor model from the Bose-Hubbard model}
In this appendix we show the steps to get to the quantum rotor model from Bose-Hubbard model. The important point is neglecting the amplitude fluctuations and assuming that $n^{\alpha}_{i} \sim n_{0}$. For clarification we use this approximation in Eq.~(\ref{eq:1}),
\begin{align}\label{eq:27}
H_{1}=&
	-\sum_{\langle ij \rangle}t_{ij}b^{\dagger A}_{i}b^{B}_{j} +H.c.
\nonumber\\=&
	-\sum_{\langle ij \rangle}\bigg( t_{ij}\sqrt{n^{A}_{i}n^{B}_{j}}e^{i(\theta^{A}_{i}-\theta^{B}_{j})} +\sqrt{n^{A}_{i}n^{B}_{j}}e^{i(\theta^{B}_{j}-\theta^{A}_{i})}\bigg)
\nonumber\\\simeq&
		2\sum_{\langle ij \rangle}t_{ij}\sqrt{n_{0}{n_{0}}}[cos(\theta^{A}_{i}-\theta^{B}_{j})]
\nonumber\\\simeq&
		2\sum_{\langle ij }n_{0} t [cos(\theta^{A}_{i}-\theta^{B}_{j})].
\end{align}

\section{Coulomb interaction and Plasma Mode}
In this appendix we show that including long range Coulomb interaction, the BAG mode explained in Sec. II and III in Eq. (\ref{eq:13}) and (\ref{eq:21}) as acoustic mode, becomes gapped plasma mode near $\Gamma$ point. The grains in the honeycomb lattice have finite charge due to large number of cooper pairs. Hence, practically we should include long range Coulomb interaction between them. Considering all the approximations discussed in Sec. II, the Hamiltonian in Eq. (\ref{eq:6}) should be modified by adding a term like $\sum_{ij \alpha \beta} U'_{ij}n_{i}^{\alpha}n_{j}^{\beta}$. Here sum on $i,j$ extends over the sites and $\alpha,\beta$ extends over the sub-lattice indices $A,B$ . $U'_{ij}$ is the bare Coulomb interaction between the grains. In order to progress further and see its effect on the collective modes, we need to incorporate some approximations when taking the Fourier transform of the Hamiltonian.
\begin{align}\label{eq:28}
\mathcal{H}'=& \sum_{{\bf k}\alpha} \biggl\{ J \Bigl(3\theta^{\alpha}_{{\bf k}}\theta^{\alpha}_{-{\bf k}}
			-
			\gamma_{\bf k}\theta^{A}_{{\bf k}}\theta^{B}_{-{\bf k}}
			-
			\gamma_{-{\bf k}}\theta^{A}_{-{\bf k}}\theta^{B}_{{\bf k}}
		\Bigr) + Un^{\alpha}_{{\bf k}}n^{\alpha}_{-{\bf k}}
\nonumber\\&  + V'_{\bm{k}} n^{A}_{{\bf k}}n^{B}_{-{\bf k}}+	V'_{\bm{k}}n^{A}_{-{\bf k}}n^{B}_{{\bf k}} + U'_{\bm{k}}n^{A}_{-{\bf k}}n^{A}_{{\bf k}}	+ U'_{\bm{k}}n^{B}_{-{\bf k}}n^{B}_{{\bf k}}) \biggr\}
\end{align}
In the above Eq. (\ref{eq:28}) $U'_{\bm{k}}$ and $V'_{\bm{k}}$ are bare Coulomb interactions. In low momentum {\it{i.e.}}, long wavelength limit, we should expect no difference in the interaction between same same sub-lattice and different sublattice.

 \begin{figure}[htb!]
  \centering
  \includegraphics[height=1.8 in ,width=3.5 in]{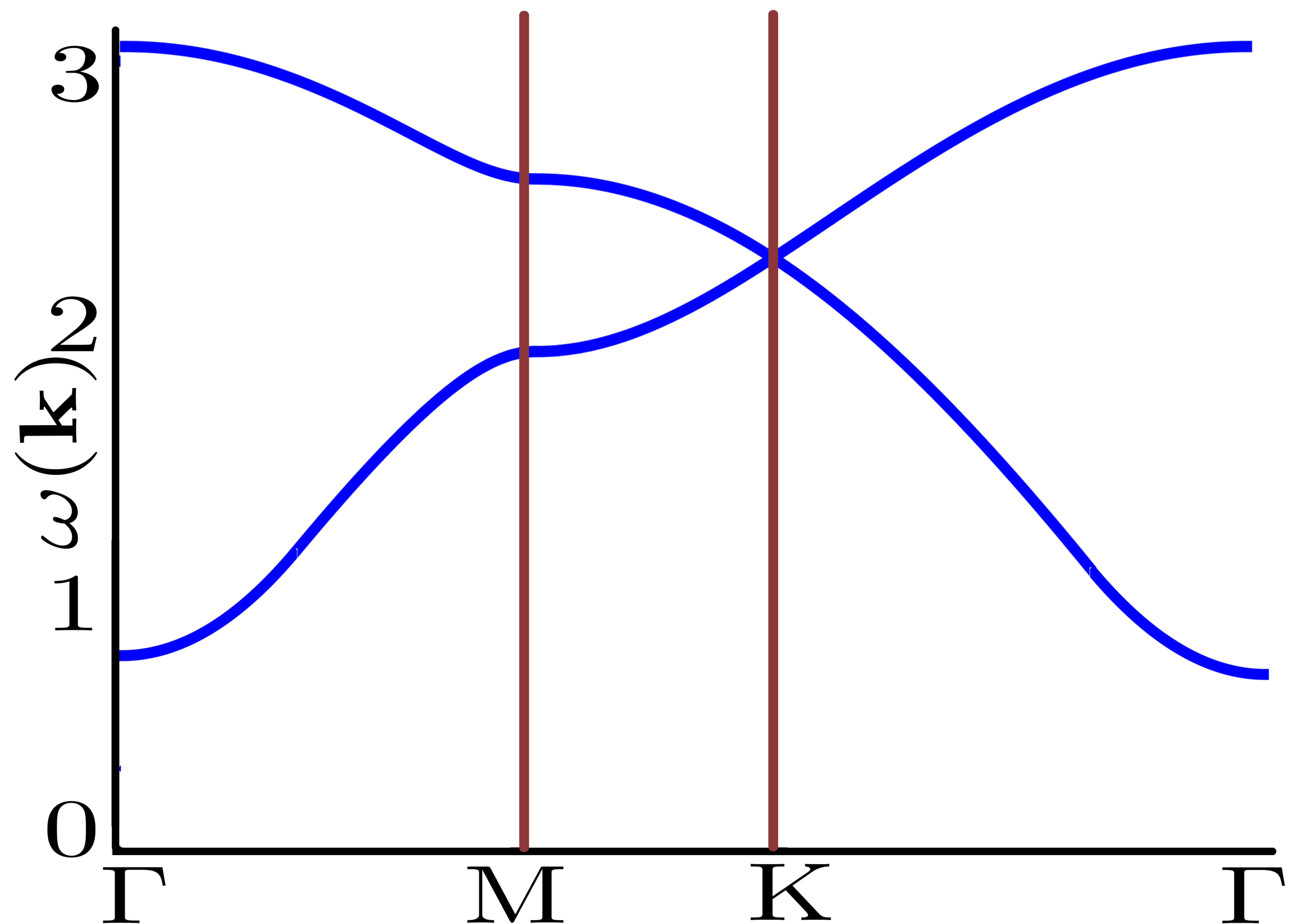}
  \caption{(Color Online) The spectra for bosonic collective modes (in units of $\sqrt{JU}$)  $\omega_{1}({\bf k})$ (leggett, top band) and $\omega_{2}({\bf k})$ (plasma, down band)  as traversed from high symmetry points $\Gamma $ to $M$ to $K$ to $\Gamma $ including the long range Coulomb interaction $U'_{ij}$. Plasma mode gap is $\sim \left( C_{1}-C_{2} \right)$ near the $\Gamma$ point.}
  \label{fig:Com_1}
 \end{figure}
 
  Therefore, in the Hamiltonian (Eq. (\ref{eq:28})),  we approximate the factors coming from summation over neighbouring vectors in the interaction terms and take them to be $ U'_{\bm{k}}$ and  $V'_{\bm{k}}$ for two sub-lattice. We expect this to be a good approximation to describe the modes near the $\Gamma$ point.  We write down the Hamiltonian in matrix notation for simplicity. We define the spinor notation ${\bm{\eta}_{k}}$ and ${\bf{\Theta}}_{\bf{k}}$ for following matrices.

\begin{equation} \label{eq:29}
 {\bm{\eta}}_{\bm{k}} = \left(n^{A}_{\bm{k}} \quad n^{B}_{\bm{k}} \right)^{T} ,\quad {\bm{\Theta}}_{\bm{k}} = \left(\theta^{A}_{\bm{k}} \quad \theta^{B}_{\bm{k}} \right)^{T}
\end{equation}\\

In terms of this matrix notation the Hamiltonian becomes ($\gamma^{R}_{\bm{k}}$ and $\gamma^{I}_{\bm{k}}$ denotes the real and imaginary part of $\gamma_{\bf{k}}$ and $\mathbb{I}$ denotes the identity matrix and $\tau$ are the Pauli matrices)
\begin{align}\label{eq:30}
\mathcal{H}= &  {{\bm{\eta}}_{\bm{k}}}^{T} \left( (U+U'_{\bm{k}})\mathbb{I} +  V'_{\bm{k}} \tau_{1} \right) {\bm{\eta}}_{-\bm{k}}
\nonumber\\  + & {{\bm{\Theta}}_{\bm{k}}}^{T} \left( 3J\mathbb{I} +J\gamma^{R}_{\bm{k}} \tau_{1} +J\gamma^{I}_{\bm{k}}\tau_{2} \right) {{\bm{\Theta}}_{-\bm{k}}}
\end{align}
We find the equations of motion for the spinors from the Hamiltonian in Eq. (\ref{eq:30}), couple them and write down the equation for the modes as,

 \begin{equation}\label{eq:31}
\ddot{\bm{\Theta}}_{\bm{k}} = -J \left(  (U+U'_{\bm{k}})\mathbb{I} +  V'_{\bm{k}} \tau_{1} \right) \left( 3J\mathbb{I} +J\gamma^{R}_{\bm{k}} \tau_{1} +J\gamma^{I}_{\bm{k}}\tau_{2} \right) {{\bm{\Theta}}}_{\bm{k}}
 \end{equation}
We will be solving this equation near the $\Gamma$ point and neglect the $\gamma^{I}_{\bm{k}}$ near the $\Gamma$ point. Now, we do an unitary transformation $\mathcal{U}=e^{\:i\frac{\pi}{4}\tau_{2}}$ to change the basis of Pauli matrices. After performing the transformation, we find
 \begin{equation}\label{eq:32}
\ddot{{\bm{\Theta}}}_{\bm{k}} = -J \left(  (U+U'_{\bm{k}})\mathbb{I} +  V'_{\bm{k}} \tau_{3} \right) \left( 3J\mathbb{I} +J\gamma^{R}_{\bm{k}} \tau_{3}  \right) {{\bm{\Theta}}}_{\bm{k}}
 \end{equation}
We take the form of the bare Coulomb interaction $U'_{\bm{k}} \sim \frac{C_{1}}{|\bm{k}|^{2}}$ and $V'_{\bm{k}} \sim \frac{C_{2}}{|\bm{k}|^{2}}$  we get the plasma mode frequency, ($C_{1}$ and $C_{2}$ are constants and we crudely assume $C_{1}>C_{2}$)
 \begin{align}\label{eq:33}
 {\omega^{2}_{2}(\bm{k}}) & \sim J(3-\gamma^{R}_{\bm{k}}) (U+U'_{\bm{k}}-V'_{\bm{k}})
 \nonumber\\  {\omega^{2}_{2}(\bm{k}})   & \sim J(\frac{3}{2}{|\bm{k}|}^{2}) \left(U+\frac{C_{1}-C_{2}}{|\bm{k}|^{2}} \right)
 \nonumber\\  {\omega^{2}_{2}(\bm{k}})   &  \sim (C_{1}-C_{2}) + JU\frac{3}{2}|\bm{k}|^{2}
 \end{align}
Therefore, we see that Coulomb interaction in our model Eq. (\ref{eq:28}) gives plasma mode in similarity with plasma mode in two band superconductor \cite{Shara}. When $C_{1}=C_{2}$ this mode becomes acoustic as explained in the main text [Eq. (\ref{eq:13}) and Eq. (\ref{eq:21})]. Including the interaction as explained in the beginning of this appendix, we expect the qualitative change of the spectra as traversed along the high symmetry points in the Brillouin zone in Fig. \ref{fig:Com_1}.

\bibliographystyle{apsrev}
\bibliography{Multimode}

\begin{thebibliography}{33}
\expandafter\ifx\csname natexlab\endcsname\relax\def\natexlab#1{#1}\fi
\expandafter\ifx\csname bibnamefont\endcsname\relax
  \def\bibnamefont#1{#1}\fi
\expandafter\ifx\csname bibfnamefont\endcsname\relax
  \def\bibfnamefont#1{#1}\fi
\expandafter\ifx\csname citenamefont\endcsname\relax
  \def\citenamefont#1{#1}\fi
\expandafter\ifx\csname url\endcsname\relax
  \def\url#1{\texttt{#1}}\fi
\expandafter\ifx\csname urlprefix\endcsname\relax\def\urlprefix{URL }\fi
\providecommand{\bibinfo}[2]{#2}
\providecommand{\eprint}[2][]{\url{#2}}

\bibitem[{\citenamefont{Castro~Neto et~al.}(2009)\citenamefont{Castro~Neto,
  Guinea, Peres, Novoselov, and Geim}}]{Castro}
\bibinfo{author}{\bibfnamefont{A.~H.} \bibnamefont{Castro~Neto}},
  \bibinfo{author}{\bibfnamefont{F.}~\bibnamefont{Guinea}},
  \bibinfo{author}{\bibfnamefont{N.~M.~R.} \bibnamefont{Peres}},
  \bibinfo{author}{\bibfnamefont{K.~S.} \bibnamefont{Novoselov}},
  \bibnamefont{and} \bibinfo{author}{\bibfnamefont{A.~K.} \bibnamefont{Geim}},
  \bibinfo{journal}{Rev. Mod. Phys.} \textbf{\bibinfo{volume}{81}},
  \bibinfo{pages}{109} (\bibinfo{year}{2009}),
  \urlprefix\url{http://link.aps.org/doi/10.1103/RevModPhys.81.109}.

\bibitem[{\citenamefont{Wehling et~al.}(2014)\citenamefont{Wehling,
  Black-Schaffer, and Balatsky}}]{Wehling}
\bibinfo{author}{\bibfnamefont{T.}~\bibnamefont{Wehling}},
  \bibinfo{author}{\bibfnamefont{A.}~\bibnamefont{Black-Schaffer}},
  \bibnamefont{and} \bibinfo{author}{\bibfnamefont{A.}~\bibnamefont{Balatsky}},
  \bibinfo{journal}{Advances in Physics} \textbf{\bibinfo{volume}{63}},
  \bibinfo{pages}{1} (\bibinfo{year}{2014}),
  \eprint{http://dx.doi.org/10.1080/00018732.2014.927109},
  \urlprefix\url{http://dx.doi.org/10.1080/00018732.2014.927109}.

\bibitem[{\citenamefont{Vafek and Vishwanath}(2014)}]{Vafek}
\bibinfo{author}{\bibfnamefont{O.}~\bibnamefont{Vafek}} \bibnamefont{and}
  \bibinfo{author}{\bibfnamefont{A.}~\bibnamefont{Vishwanath}},
  \bibinfo{journal}{Annual Review of Condensed Matter Physics}
  \textbf{\bibinfo{volume}{5}}, \bibinfo{pages}{83} (\bibinfo{year}{2014}),
  \eprint{http://dx.doi.org/10.1146/annurev-conmatphys-031113-133841},
  \urlprefix\url{http://dx.doi.org/10.1146/annurev-conmatphys-031113-133841}.

\bibitem[{\citenamefont{Semenoff}(1984)}]{Semenoff}
\bibinfo{author}{\bibfnamefont{G.~W.} \bibnamefont{Semenoff}},
  \bibinfo{journal}{Phys. Rev. Lett.} \textbf{\bibinfo{volume}{53}},
  \bibinfo{pages}{2449} (\bibinfo{year}{1984}),
  \urlprefix\url{http://link.aps.org/doi/10.1103/PhysRevLett.53.2449}.

\bibitem[{\citenamefont{Park and Choi}(2015)}]{Park}
\bibinfo{author}{\bibfnamefont{J.-S.} \bibnamefont{Park}} \bibnamefont{and}
  \bibinfo{author}{\bibfnamefont{H.~J.} \bibnamefont{Choi}},
  \bibinfo{journal}{Phys. Rev. B} \textbf{\bibinfo{volume}{92}},
  \bibinfo{pages}{045402} (\bibinfo{year}{2015}),
  \urlprefix\url{http://link.aps.org/doi/10.1103/PhysRevB.92.045402}.

\bibitem[{\citenamefont{Tarruell et~al.}(2012)\citenamefont{Tarruell, Greif,
  Uehlinger, Shvets, Jotzu, and Esslinger}}]{Tarr}
\bibinfo{author}{\bibfnamefont{L.}~\bibnamefont{Tarruell}},
  \bibinfo{author}{\bibfnamefont{D.}~\bibnamefont{Greif}},
  \bibinfo{author}{\bibfnamefont{T.}~\bibnamefont{Uehlinger}},
  \bibinfo{author}{\bibnamefont{Shvets}},
  \bibinfo{author}{\bibfnamefont{G.}~\bibnamefont{Jotzu}}, \bibnamefont{and}
  \bibinfo{author}{\bibfnamefont{T.}~\bibnamefont{Esslinger}},
  \bibinfo{journal}{Nature} \textbf{\bibinfo{volume}{483}},
  \bibinfo{pages}{302} (\bibinfo{year}{2012}),
  \urlprefix\url{http://dx.doi.org/10.1038/nature10871}.

\bibitem[{\citenamefont{Gomes et~al.}(2012)\citenamefont{Gomes, Mar, Ko,
  Guinea, and Manoharan}}]{Man1}
\bibinfo{author}{\bibfnamefont{K.~K.} \bibnamefont{Gomes}},
  \bibinfo{author}{\bibfnamefont{W.}~\bibnamefont{Mar}},
  \bibinfo{author}{\bibfnamefont{W.}~\bibnamefont{Ko}},
  \bibinfo{author}{\bibfnamefont{F.}~\bibnamefont{Guinea}}, \bibnamefont{and}
  \bibinfo{author}{\bibfnamefont{H.~C.} \bibnamefont{Manoharan}},
  \bibinfo{journal}{Nature} \textbf{\bibinfo{volume}{483}},
  \bibinfo{pages}{306} (\bibinfo{year}{2012}),
  \urlprefix\url{http://dx.doi.org/10.1038/nature10941}.

\bibitem[{\citenamefont{Marco et~al.}(2013)\citenamefont{Marco, Francisco,
  Maciej, Hari, and Vittorio}}]{Man2}
\bibinfo{author}{\bibfnamefont{P.}~\bibnamefont{Marco}},
  \bibinfo{author}{\bibfnamefont{G.}~\bibnamefont{Francisco}},
  \bibinfo{author}{\bibfnamefont{L.}~\bibnamefont{Maciej}},
  \bibinfo{author}{\bibfnamefont{C.~M.} \bibnamefont{Hari}}, \bibnamefont{and}
  \bibinfo{author}{\bibfnamefont{P.}~\bibnamefont{Vittorio}},
  \bibinfo{journal}{Nat Nano} \textbf{\bibinfo{volume}{8}},
  \bibinfo{pages}{625} (\bibinfo{year}{2013}),
  \urlprefix\url{doi:10.1038/nnano.2013.161}.

\bibitem[{\citenamefont{Hammar et~al.}(2013)\citenamefont{Hammar, Berggren, and
  Fransson}}]{Hammar}
\bibinfo{author}{\bibfnamefont{H.}~\bibnamefont{Hammar}},
  \bibinfo{author}{\bibfnamefont{P.}~\bibnamefont{Berggren}}, \bibnamefont{and}
  \bibinfo{author}{\bibfnamefont{J.}~\bibnamefont{Fransson}},
  \bibinfo{journal}{Phys. Rev. B} \textbf{\bibinfo{volume}{88}},
  \bibinfo{pages}{245418} (\bibinfo{year}{2013}),
  \urlprefix\url{http://link.aps.org/doi/10.1103/PhysRevB.88.245418}.

\bibitem[{\citenamefont{Weick et~al.}(2013)\citenamefont{Weick, Woollacott,
  Barnes, Hess, and Mariani}}]{Weick}
\bibinfo{author}{\bibfnamefont{G.}~\bibnamefont{Weick}},
  \bibinfo{author}{\bibfnamefont{C.}~\bibnamefont{Woollacott}},
  \bibinfo{author}{\bibfnamefont{W.~L.} \bibnamefont{Barnes}},
  \bibinfo{author}{\bibfnamefont{O.}~\bibnamefont{Hess}}, \bibnamefont{and}
  \bibinfo{author}{\bibfnamefont{E.}~\bibnamefont{Mariani}},
  \bibinfo{journal}{Phys. Rev. Lett.} \textbf{\bibinfo{volume}{110}},
  \bibinfo{pages}{106801} (\bibinfo{year}{2013}),
  \urlprefix\url{http://link.aps.org/doi/10.1103/PhysRevLett.110.106801}.

\bibitem[{\citenamefont{Chen and Wu}(2011)}]{Chen}
\bibinfo{author}{\bibfnamefont{Z.}~\bibnamefont{Chen}} \bibnamefont{and}
  \bibinfo{author}{\bibfnamefont{B.}~\bibnamefont{Wu}}, \bibinfo{journal}{Phys.
  Rev. Lett.} \textbf{\bibinfo{volume}{107}}, \bibinfo{pages}{065301}
  (\bibinfo{year}{2011}),
  \urlprefix\url{http://link.aps.org/doi/10.1103/PhysRevLett.107.065301}.

\bibitem[{\citenamefont{Wang}(2010)}]{Wang}
\bibinfo{author}{\bibfnamefont{F.}~\bibnamefont{Wang}}, \bibinfo{journal}{Phys.
  Rev. B} \textbf{\bibinfo{volume}{82}}, \bibinfo{pages}{024419}
  (\bibinfo{year}{2010}),
  \urlprefix\url{http://link.aps.org/doi/10.1103/PhysRevB.82.024419}.

\bibitem[{\citenamefont{Ma et~al.}(2015)\citenamefont{Ma, Khanikaev, Mousavi,
  and Shvets}}]{Khan}
\bibinfo{author}{\bibfnamefont{T.}~\bibnamefont{Ma}},
  \bibinfo{author}{\bibfnamefont{A.~B.} \bibnamefont{Khanikaev}},
  \bibinfo{author}{\bibfnamefont{S.~H.} \bibnamefont{Mousavi}},
  \bibnamefont{and} \bibinfo{author}{\bibfnamefont{G.}~\bibnamefont{Shvets}},
  \bibinfo{journal}{Phys. Rev. Lett.} \textbf{\bibinfo{volume}{114}},
  \bibinfo{pages}{127401} (\bibinfo{year}{2015}),
  \urlprefix\url{http://link.aps.org/doi/10.1103/PhysRevLett.114.127401}.

\bibitem[{\citenamefont{Haddad and Carr}(2009)}]{Haddad}
\bibinfo{author}{\bibfnamefont{L.}~\bibnamefont{Haddad}} \bibnamefont{and}
  \bibinfo{author}{\bibfnamefont{L.}~\bibnamefont{Carr}},
  \bibinfo{journal}{Physica D: Nonlinear Phenomena}
  \textbf{\bibinfo{volume}{238}}, \bibinfo{pages}{1413 }
  (\bibinfo{year}{2009}), ISSN \bibinfo{issn}{0167-2789},
  \bibinfo{note}{nonlinear Phenomena in Degenerate Quantum Gases},
  \urlprefix\url{http://www.sciencedirect.com/science/article/pii/S0167278909000372}.

\bibitem[{\citenamefont{Bellec et~al.}(2013)\citenamefont{Bellec, Kuhl,
  Montambaux, and Mortessagne}}]{Bellec}
\bibinfo{author}{\bibfnamefont{M.}~\bibnamefont{Bellec}},
  \bibinfo{author}{\bibfnamefont{U.}~\bibnamefont{Kuhl}},
  \bibinfo{author}{\bibfnamefont{G.}~\bibnamefont{Montambaux}},
  \bibnamefont{and}
  \bibinfo{author}{\bibfnamefont{F.}~\bibnamefont{Mortessagne}},
  \bibinfo{journal}{Phys. Rev. B} \textbf{\bibinfo{volume}{88}},
  \bibinfo{pages}{115437} (\bibinfo{year}{2013}),
  \urlprefix\url{http://link.aps.org/doi/10.1103/PhysRevB.88.115437}.

\bibitem[{\citenamefont{Jacqmin et~al.}(2014)\citenamefont{Jacqmin, Carusotto,
  Sagnes, Abbarchi, Solnyshkov, Malpuech, Galopin, Lema\^{\i}tre, Bloch, and
  Amo}}]{Jacqmin}
\bibinfo{author}{\bibfnamefont{T.}~\bibnamefont{Jacqmin}},
  \bibinfo{author}{\bibfnamefont{I.}~\bibnamefont{Carusotto}},
  \bibinfo{author}{\bibfnamefont{I.}~\bibnamefont{Sagnes}},
  \bibinfo{author}{\bibfnamefont{M.}~\bibnamefont{Abbarchi}},
  \bibinfo{author}{\bibfnamefont{D.~D.} \bibnamefont{Solnyshkov}},
  \bibinfo{author}{\bibfnamefont{G.}~\bibnamefont{Malpuech}},
  \bibinfo{author}{\bibfnamefont{E.}~\bibnamefont{Galopin}},
  \bibinfo{author}{\bibfnamefont{A.}~\bibnamefont{Lema\^{\i}tre}},
  \bibinfo{author}{\bibfnamefont{J.}~\bibnamefont{Bloch}}, \bibnamefont{and}
  \bibinfo{author}{\bibfnamefont{A.}~\bibnamefont{Amo}},
  \bibinfo{journal}{Phys. Rev. Lett.} \textbf{\bibinfo{volume}{112}},
  \bibinfo{pages}{116402} (\bibinfo{year}{2014}),
  \urlprefix\url{http://link.aps.org/doi/10.1103/PhysRevLett.112.116402}.

\bibitem[{\citenamefont{Duca et~al.}(2015)\citenamefont{Duca, Li, Reitter,
  Bloch, Schleier-Smith, and Schneider}}]{Duca}
\bibinfo{author}{\bibfnamefont{L.}~\bibnamefont{Duca}},
  \bibinfo{author}{\bibfnamefont{T.}~\bibnamefont{Li}},
  \bibinfo{author}{\bibfnamefont{M.}~\bibnamefont{Reitter}},
  \bibinfo{author}{\bibfnamefont{I.}~\bibnamefont{Bloch}},
  \bibinfo{author}{\bibfnamefont{M.}~\bibnamefont{Schleier-Smith}},
  \bibnamefont{and}
  \bibinfo{author}{\bibfnamefont{U.}~\bibnamefont{Schneider}},
  \bibinfo{journal}{Science} \textbf{\bibinfo{volume}{347}},
  \bibinfo{pages}{288} (\bibinfo{year}{2015}), ISSN \bibinfo{issn}{0036-8075},
  \eprint{http://science.sciencemag.org/content/347/6219/288.full.pdf},
  \urlprefix\url{http://science.sciencemag.org/content/347/6219/288}.

\bibitem[{\citenamefont{Li et~al.}(2015)\citenamefont{Li, Sengupta, Batrouni,
  Miniatura, and Gr\'emaud}}]{Sengupta}
\bibinfo{author}{\bibfnamefont{Y.}~\bibnamefont{Li}},
  \bibinfo{author}{\bibfnamefont{P.}~\bibnamefont{Sengupta}},
  \bibinfo{author}{\bibfnamefont{G.~G.} \bibnamefont{Batrouni}},
  \bibinfo{author}{\bibfnamefont{C.}~\bibnamefont{Miniatura}},
  \bibnamefont{and}
  \bibinfo{author}{\bibfnamefont{B.}~\bibnamefont{Gr\'emaud}},
  \bibinfo{journal}{Phys. Rev. A} \textbf{\bibinfo{volume}{92}},
  \bibinfo{pages}{043605} (\bibinfo{year}{2015}),
  \urlprefix\url{http://link.aps.org/doi/10.1103/PhysRevA.92.043605}.

\bibitem[{\citenamefont{Chern and Saxena}(2015)}]{Chern}
\bibinfo{author}{\bibfnamefont{G.-W.} \bibnamefont{Chern}} \bibnamefont{and}
  \bibinfo{author}{\bibfnamefont{A.}~\bibnamefont{Saxena}},
  \bibinfo{journal}{Opt. Lett.} \textbf{\bibinfo{volume}{40}},
  \bibinfo{pages}{5806} (\bibinfo{year}{2015}),
  \urlprefix\url{http://ol.osa.org/abstract.cfm?URI=ol-40-24-5806}.

\bibitem[{\citenamefont{Fransson et~al.}(2015)\citenamefont{Fransson,
  Black-Schaffer, and Balatsky}}]{Jonas}
\bibinfo{author}{\bibfnamefont{J.}~\bibnamefont{Fransson}},
  \bibinfo{author}{\bibfnamefont{A.~M.} \bibnamefont{Black-Schaffer}},
  \bibnamefont{and} \bibinfo{author}{\bibfnamefont{A.}~\bibnamefont{Balatsky}},
  \bibinfo{journal}{preprint, to be published}  (\bibinfo{year}{2015}),
  \urlprefix\url{http://arxiv.org/abs/1512.04902}.

\bibitem[{\citenamefont{Beloborodov et~al.}(2007)\citenamefont{Beloborodov,
  Lopatin, Vinokur, and Efetov}}]{Belo}
\bibinfo{author}{\bibfnamefont{I.~S.} \bibnamefont{Beloborodov}},
  \bibinfo{author}{\bibfnamefont{A.~V.} \bibnamefont{Lopatin}},
  \bibinfo{author}{\bibfnamefont{V.~M.} \bibnamefont{Vinokur}},
  \bibnamefont{and} \bibinfo{author}{\bibfnamefont{K.~B.}
  \bibnamefont{Efetov}}, \bibinfo{journal}{Rev. Mod. Phys.}
  \textbf{\bibinfo{volume}{79}}, \bibinfo{pages}{469} (\bibinfo{year}{2007}),
  \urlprefix\url{http://link.aps.org/doi/10.1103/RevModPhys.79.469}.

\bibitem[{\citenamefont{Krull et~al.}(2015)\citenamefont{Krull, Bittner, Uhrig,
  Manske, and Schnyder}}]{Krull}
\bibinfo{author}{\bibfnamefont{H.}~\bibnamefont{Krull}},
  \bibinfo{author}{\bibfnamefont{N.}~\bibnamefont{Bittner}},
  \bibinfo{author}{\bibfnamefont{G.~S.} \bibnamefont{Uhrig}},
  \bibinfo{author}{\bibfnamefont{D.}~\bibnamefont{Manske}}, \bibnamefont{and}
  \bibinfo{author}{\bibfnamefont{A.~P.} \bibnamefont{Schnyder}},
  \bibinfo{journal}{preprint, to be published}  (\bibinfo{year}{2015}),
  \urlprefix\url{http://arxiv.org/abs/1512.08121}.

\bibitem[{\citenamefont{Gan et~al.}(2007)\citenamefont{Gan, Wen, Ye, Li, Yang,
  and Yu}}]{Gan}
\bibinfo{author}{\bibfnamefont{J.~Y.} \bibnamefont{Gan}},
  \bibinfo{author}{\bibfnamefont{Y.~C.} \bibnamefont{Wen}},
  \bibinfo{author}{\bibfnamefont{J.}~\bibnamefont{Ye}},
  \bibinfo{author}{\bibfnamefont{T.}~\bibnamefont{Li}},
  \bibinfo{author}{\bibfnamefont{S.-J.} \bibnamefont{Yang}}, \bibnamefont{and}
  \bibinfo{author}{\bibfnamefont{Y.}~\bibnamefont{Yu}}, \bibinfo{journal}{Phys.
  Rev. B} \textbf{\bibinfo{volume}{75}}, \bibinfo{pages}{214509}
  (\bibinfo{year}{2007}),
  \urlprefix\url{http://link.aps.org/doi/10.1103/PhysRevB.75.214509}.

\bibitem[{\citenamefont{Fisher et~al.}(1989)\citenamefont{Fisher, Weichman,
  Grinstein, and Fisher}}]{Fisher}
\bibinfo{author}{\bibfnamefont{M.~P.~A.} \bibnamefont{Fisher}},
  \bibinfo{author}{\bibfnamefont{P.~B.} \bibnamefont{Weichman}},
  \bibinfo{author}{\bibfnamefont{G.}~\bibnamefont{Grinstein}},
  \bibnamefont{and} \bibinfo{author}{\bibfnamefont{D.~S.}
  \bibnamefont{Fisher}}, \bibinfo{journal}{Phys. Rev. B}
  \textbf{\bibinfo{volume}{40}}, \bibinfo{pages}{546} (\bibinfo{year}{1989}),
  \urlprefix\url{http://link.aps.org/doi/10.1103/PhysRevB.40.546}.

\bibitem[{\citenamefont{Kisker and Rieger}(1997)}]{Kisker}
\bibinfo{author}{\bibfnamefont{J.}~\bibnamefont{Kisker}} \bibnamefont{and}
  \bibinfo{author}{\bibfnamefont{H.}~\bibnamefont{Rieger}},
  \bibinfo{journal}{Phys. Rev. B} \textbf{\bibinfo{volume}{55}},
  \bibinfo{pages}{R11981} (\bibinfo{year}{1997}),
  \urlprefix\url{http://link.aps.org/doi/10.1103/PhysRevB.55.R11981}.

\bibitem[{\citenamefont{Das and Doniach}(1999)}]{Das}
\bibinfo{author}{\bibfnamefont{D.}~\bibnamefont{Das}} \bibnamefont{and}
  \bibinfo{author}{\bibfnamefont{S.}~\bibnamefont{Doniach}},
  \bibinfo{journal}{Phys. Rev. B} \textbf{\bibinfo{volume}{60}},
  \bibinfo{pages}{1261} (\bibinfo{year}{1999}),
  \urlprefix\url{http://link.aps.org/doi/10.1103/PhysRevB.60.1261}.

\bibitem[{\citenamefont{Cha et~al.}(1991)\citenamefont{Cha, Fisher, Girvin,
  Wallin, and Young}}]{Cha}
\bibinfo{author}{\bibfnamefont{M.-C.} \bibnamefont{Cha}},
  \bibinfo{author}{\bibfnamefont{M.~P.~A.} \bibnamefont{Fisher}},
  \bibinfo{author}{\bibfnamefont{S.~M.} \bibnamefont{Girvin}},
  \bibinfo{author}{\bibfnamefont{M.}~\bibnamefont{Wallin}}, \bibnamefont{and}
  \bibinfo{author}{\bibfnamefont{A.~P.} \bibnamefont{Young}},
  \bibinfo{journal}{Phys. Rev. B} \textbf{\bibinfo{volume}{44}},
  \bibinfo{pages}{6883} (\bibinfo{year}{1991}),
  \urlprefix\url{http://link.aps.org/doi/10.1103/PhysRevB.44.6883}.

\bibitem[{\citenamefont{Sharapov et~al.}(2002)\citenamefont{Sharapov, Gusynin,
  and Beck}}]{Shara}
\bibinfo{author}{\bibfnamefont{S.}~\bibnamefont{Sharapov}},
  \bibinfo{author}{\bibfnamefont{V.}~\bibnamefont{Gusynin}}, \bibnamefont{and}
  \bibinfo{author}{\bibfnamefont{H.}~\bibnamefont{Beck}}, \bibinfo{journal}{The
  European Physical Journal B - Condensed Matter and Complex Systems}
  \textbf{\bibinfo{volume}{30}}, \bibinfo{pages}{45} (\bibinfo{year}{2002}),
  ISSN \bibinfo{issn}{1434-6028},
  \urlprefix\url{http://dx.doi.org/10.1140/epjb/e2002-00356-9}.

\bibitem[{\citenamefont{Lin and Hu}(2012)}]{Lin}
\bibinfo{author}{\bibfnamefont{S.-Z.} \bibnamefont{Lin}} \bibnamefont{and}
  \bibinfo{author}{\bibfnamefont{X.}~\bibnamefont{Hu}}, \bibinfo{journal}{Phys.
  Rev. Lett.} \textbf{\bibinfo{volume}{108}}, \bibinfo{pages}{177005}
  (\bibinfo{year}{2012}),
  \urlprefix\url{http://link.aps.org/doi/10.1103/PhysRevLett.108.177005}.

\bibitem[{\citenamefont{Hatsugai et~al.}(2013)\citenamefont{Hatsugai, Morimoto,
  Kawarabayashi, Hamamoto, and Aoki}}]{Hatsu}
\bibinfo{author}{\bibfnamefont{Y.}~\bibnamefont{Hatsugai}},
  \bibinfo{author}{\bibfnamefont{T.}~\bibnamefont{Morimoto}},
  \bibinfo{author}{\bibfnamefont{T.}~\bibnamefont{Kawarabayashi}},
  \bibinfo{author}{\bibfnamefont{Y.}~\bibnamefont{Hamamoto}}, \bibnamefont{and}
  \bibinfo{author}{\bibfnamefont{H.}~\bibnamefont{Aoki}}, \bibinfo{journal}{New
  Journal of Physics} \textbf{\bibinfo{volume}{15}}, \bibinfo{pages}{035023}
  (\bibinfo{year}{2013}),
  \urlprefix\url{http://stacks.iop.org/1367-2630/15/i=3/a=035023}.

\bibitem[{\citenamefont{Pethick and Smith}(2004)}]{Pethick}
\bibinfo{editor}{\bibfnamefont{C.}~\bibnamefont{Pethick}} \bibnamefont{and}
  \bibinfo{editor}{\bibfnamefont{H.}~\bibnamefont{Smith}}, eds.,
  \emph{\bibinfo{title}{Bose Einstein Condensation in Dilute Gases}}
  (\bibinfo{publisher}{Cambridge University Press},
  \bibinfo{address}{Cambridge}, \bibinfo{year}{2004}).

\bibitem[{\citenamefont{Wang et~al.}(2015)\citenamefont{Wang, Deng, Liu, and
  Liu}}]{WangJ}
\bibinfo{author}{\bibfnamefont{J.}~\bibnamefont{Wang}},
  \bibinfo{author}{\bibfnamefont{S.}~\bibnamefont{Deng}},
  \bibinfo{author}{\bibfnamefont{Z.}~\bibnamefont{Liu}}, \bibnamefont{and}
  \bibinfo{author}{\bibfnamefont{Z.}~\bibnamefont{Liu}},
  \bibinfo{journal}{National Science Review} \textbf{\bibinfo{volume}{2}},
  \bibinfo{pages}{22} (\bibinfo{year}{2015}),
  \urlprefix\url{http://nsr.oxfordjournals.org/content/2/1/22.abstract}.

\bibitem[{\citenamefont{Zhao and Paramekanti}(2006)}]{Zhao}
\bibinfo{author}{\bibfnamefont{E.}~\bibnamefont{Zhao}} \bibnamefont{and}
  \bibinfo{author}{\bibfnamefont{A.}~\bibnamefont{Paramekanti}},
  \bibinfo{journal}{Phys. Rev. Lett.} \textbf{\bibinfo{volume}{97}},
  \bibinfo{pages}{230404} (\bibinfo{year}{2006}),
  \urlprefix\url{http://link.aps.org/doi/10.1103/PhysRevLett.97.230404}.

\end{thebibliography}

\end{document}